# Tangential and normal partial slip at the liquid-fluid interfaces: An application to a small liquid droplet, gas bubble, and aerosol

Peter Lebedev-Stepanov

Shubnikov Institute of Crystallography, Kurchatov Complex of Crystallography and Photonics, NRC "Kurchatov Institute", Leninskiy Prospekt 59, Moscow 119333, Russia

E-mail: lebstep.p@crys.ras.ru

**Abstract.** An analytical solution is obtained for the problem of the slow movement of a small drop of a fluid in another immiscible fluid in an infinitely large reservoir with the boundary condition of the normal slip and/or tangential partial slip at the liquid-liquid or gas-liquid interface. That generalizes the conventional Navier's and Maxwellian boundary conditions of partial slip. Normal slip is accompanied by the density gradient in the fluid and is applicable only if one of the phases in contact at the interface is a gas. At the first time the generalization of the Hadamard-Rybczynski equation (HRE) is discussed within the formalism of two slip lengths. It is proven that each of the two fluids separated by an interface has its own slip length. The slip lengths of the fluids that contact each other through an interface are related by the ratio of the viscosities of these fluids. New equations describing the terminal velocity of gas bubble rise and aerosol falling have been obtained. The theoretical results are in good agreement with the experiments at moderate Knudsen numbers, Kn<0.25. It has been shown that the gas density within a rising bubble and around a falling droplet in the air is not uniform. The relative magnitude of the density increment increases with the size of the bubble or aerosol. Presumably, the best applicability of the generalized HRE should be expected for the interface of hydrophobic liquid and hydrophilic one (water – hydrocarbons, water – higher alcohols, in general: aqueous emulsions, water – lipophilic organic liquids and oils, etc.). These are quite important emulsions in practical terms, for example, for the oil industry and medicine. Experimental methods for determining the slip length are considered.

## 1. Introduction

Small spherical particles in various states of aggregation (solid, liquid, gas), which move in an external fluid under the action of the gravity, the buoyancy and the drag force, are an important object for studying the properties of substances and materials, in wide range of the applications from printing technologies and droplet microfluidics [1-2] to atmospheric research, ecology and biomedicine [3-5], as well as a testing ground for theories at the intersection of the fluid mechanics and surface science [6].

The interaction of a particle and a surrounding medium occurs at their spherical interface. If the velocities of the contacting phases at the interface differ abruptly, there is a partial or complete slip. Otherwise, if the velocities of the contacting phases are equal, there is no-slip boundary condition [7-9]. When describing slip at an interface, the nature of the phases in contact, the characteristic size of the system, the phase velocities, and the magnitude of the velocity jump in the region of interphase contact are of fundamental importance.

When discussing the slip of a liquid or gas near a solid wall, then a liquid-solid and a gas-solid interfaces are considering, respectively. The effects of a liquid slip and a gas slip near a solid wall have fundamentally different physical origins. In the first case, this is associated with mechanical friction on a two-dimensional interface, while in the second, it is associated with kinetic effects in a near-wall gas layer with a thickness on the order of the mean free path $L$. The properties of that layer, so-called Knudsen layer, differ significantly from the main volume of gas, which manifests itself in the form of a slip effect.

The Knudsen number is the ratio of the mean free path to the characteristic size of the system, which in the case under consideration is the radius $R$ of a spherical particle, Kn=$L$/R. That number determines the relative importance of the Knudsen layer in the process of momentum transfer in the gas near interface [10].

Small Knudsen numbers (Kn $\rightarrow$ 0) correspond to the continuum flow: Gas behaves as a continuous fluid. The Navier–Stokes equations with no-slip boundary conditions are applied.

Moderate Knudsen numbers (Kn < 0.25) correspond to the slip flow: The continuous fluid assumption still holds, but the gas at the wall in the Knudsen layer slips over the particle surface. In this case, the Navier–Stokes equations with the partial or full slip boundary conditions are applied.

Large Knudsen numbers (0.25 < Kn < 10) correspond to the transitional flow: Molecular and continuum effects both play major roles. The Navier–Stokes equations break down entirely, and specialized kinetic theory models are required

The largest Knudsen numbers (Kn > 10) correspond to the free molecular flow: Individual gas molecules strike and reflect off the particle independently. The kinetic equations, for example, the Boltzmann equation, are usually applied.

Interfacial momentum transfer via slip is described by a special type of boundary conditions. In the case of an interface between two incompressible fluids or a liquid and a solid, slip friction transfers only tangentially directed momentum. If one of the phases is a gas, normal slip also occurs, transferring momentum along the normal to the interface, which may be accompanied by a density gradient.

Boundary condition for tangential slip at the liquid-solid interface was introduced by Navier [11] and Basset [12] in relation to the Navier–Stokes equations. Similar condition at the gas-solid interface was introduced by J. C. Maxwell [13-14] on the basis of molecular kinetic theory and named in his honor. The application of the Maxwellian kinetic condition at the gas-liquid interface requires its precise coupling with the Navier hydrodynamic condition, i.e. translation into the terms of hydrodynamics, taking into account the difference in the definition of kinetic and hydrodynamic slip lengths and absence of normal slip in an incompressible fluid.

This problem, in such a specific form, has not yet been discussed in the scientific literature. To solve it in this paper, we develop a slip length formalism for the liquid-liquid interface, since one can introduce a slip length for each of the two contacting liquids, unlike the case of a liquid-solid interface, where only one liquid and, accordingly, only one slip length can be involved. Then we extend the two-slip length formalism to the gas-liquid interface.

The object of this paper is the isothermal motion of a small liquid droplet or gas bubble in another liquid under the influence of gravity, as well as the similar motion of a small liquid droplet in the air under normal conditions with the tangential slip and the normal slip (if applicable) at the liquid-liquid or gas-liquid interface. The liquid medium is considered incompressible, while the gaseous medium is assumed to have a density deviation from the mean value.

We use the approximation of small or moderate Knudsen numbers, for which continuous slip flow dynamics can be used for both gas and liquid. The Knudsen near-wall gas layer manifests itself as slip at the gas-liquid interface.

Droplet motion is assumed to be stationary, corresponding to a small Reynolds number $\mathrm{Re} = 2\rho R V \eta^{-1}$, where ρ is the mass density of the liquid, $\eta$ is the dynamic viscosity, and $V$ is the characteristic velocity of the hydrodynamic flow. Condition Re << 1 corresponds to the description using the Stokes equations, i.e., linearized by the stationary Navier–Stokes equations [15], and also limits the size of the droplet or gas bubble.

The small size of the droplet ensures that surface tension forces dominate the volume forces associated with both gravitational and hydrodynamic pressure. Therefore, in the model under consideration, we assume, with sufficient accuracy, that the shape of the droplet or bubble is spherical. This is guaranteed by the smallness of the Bond number Bo and capillary number Ca [16-17]: $\mathrm{Bo} = R^2 a^{-2} << 1$, where $a = (2\sigma)^{\frac{1}{2}}(\rho g)^{-\frac{1}{2}}$ is the capillary constant, and σ is the surface tension of the interface, $g$ is the acceleration due to gravity; $\mathrm{Ca} = \eta V \sigma^{-1} << 1$.

Under the above assumptions, the Hadamard–Rybczynski equation (HRE) [18-20] was derived, which describes the velocity $V$ of the steady-state isothermal motion of a spherical liquid droplet with viscosity η' and density ρ' in an external immiscible fluid with viscosity η and density ρ under the action of gravity and no-slip condition at the interface:

$$V = \frac{2(\rho - \rho')gR^2}{3\eta}\frac{1+\kappa}{2+3\kappa}, \quad \kappa = \frac{\eta'}{\eta}.$$

If the ratio of the viscosities of the droplet and the external fluid tends to infinity, i.e., $\kappa \to \infty$, one can obtain the Stokes formula

$$V = \frac{2(\rho - \rho')gR^2}{9\eta},$$

which describes the motion of a solid particle in the external fluid. If $\kappa \to \infty$, HRE takes the form

$$V = \frac{(\rho - \rho')gR^2}{3\eta},$$

which is expected to describe the rise of a small gas bubble in a liquid [20].

In an accordance with the HRE, the condition of small Reynolds number sets the following limitation on the radius of a droplet:

$$R \ll \left( \frac{\eta^2}{\rho|\rho-\rho'|g} \frac{2+3\kappa}{1+\kappa} \right)^{\frac{1}{3}}.$$

For a water droplet falling in the air under the normal conditions with $\eta' \approx 10^{-3}\mathrm{Pa \cdot s}$, $\rho' \approx 1000\mathrm{kg \cdot m^{-3}}$, $\eta \approx 1.84 \cdot 10^{-5}\mathrm{Pa \cdot s}$, $\rho \approx 1.17\mathrm{kg \cdot m^{-3}}$, the droplet radius constraint is $R \ll 45\,\mu\mathrm{m}$. For a bubble rising in a reservoir of water, the radius is determined by the inequality $R \ll 60\,\mu\mathrm{m}$. For the air under normal conditions, the mean free path is $L \approx 67.3\,\mathrm{nm}$ (air humidity changes this by no more than a few percent) [21-22]. Thus, in the Hadamard and Rybczynski model, for a water droplet falling in the air, we have $\mathrm{Kn} \gg 0.0015$, and for a bubble rising in water $\mathrm{Kn} \gg 0.001$.

According to the established consensus, the rise of small gas bubbles at $R < 60\,\mu\mathrm{m}$ is described by the HRE at $\kappa \to \infty$ [20, 23, 24], and the deviation from this equation observed in experiments is explained by the presence of the unaccounted foreign surfactants. However, if $R < 0.5\,\mu\mathrm{m}$ ($\mathrm{Kn} > 0.1$), from the above estimates it follows that the HRE, originally based on the no-slip condition that corresponds to $\mathrm{Kn} \to 0$, taking into account the range of Reynolds numbers accessible to it, cannot be used without introducing the slip corrections in the case where one of the contacting phases is a gas.

The same observation applies when describing the fall of an aerosol (a small liquid droplet or a spherical solid particle) in the atmosphere. However, in this case, over the past 100 years, many studies, both experimental and theoretical, have appeared to consider the Knudsen-number-dependent corrections for slip in the gas phase. Cunningham [25] derived slip correction factor to the Stokes formula by bridging classical hydrodynamics with kinetic theory, replacing the no-slip condition with a slip velocity that accounts for particle size relative to the mean free path by equating macroscopic drag with molecular momentum transfer. He established a following linear correction factor:

$$C_c = 1 + A\text{Kn},$$

Millikan [26-28] experimentally confirmed the validity of such a correction for Kn < 0.3. Since then, the specific value of $A$ has been calculated in various ways, empirically or by solving kinetic equations, for different specific systems under different external conditions [6,29-33].

On the other hand, the HRE has been generalized by introducing the tangential slip condition instead of the no-slip condition [34-36]. Taking into account the tangential slip at the liquid-liquid interface in the framework HRE approach brings the description of the slip process to a new level. The presence of two liquids sliding at a common interface requires an introduction of two slip lengths at once.

However, the authors of [34-36] use the generally accepted, but insufficient in this case, formalism of one slip length, by analogy with the original Navier and Basset slip condition at liquid-solid interface. This significantly limits the applicability of the generalization and indicates a desirable direction for further development of that approach.

Another problem with this generalization of the HRE is the failure to account for normal slip. This makes it incorrect to describe the gas-liquid interface.

At the same time, there is a methodological disconnect between the approach embodied in the works of Cunningham and Millican [25-28] for considering slip at the boundary between a gas medium and a dense phase and the method for generalizing the HRE by introducing slip effects into this equation, proposed in [34-36]. The approaches used to describe the fall of the aerosols in the air, the rise of the bubbles in the liquid at the moderate Knudsen number, and a sliding at the liquid-liquid interface could be considered within a general framework.

The purpose of this paper is to attempt to fill these gaps. First, the two-slip length approach should be developed to describe by the Stokes equations for an incompressible fluid.

Second, a universal generalization of the HRE is needed for its application to the gas-liquid interface – rising bubbles or falling aerosols – taking into account normal slip. This requires the development of a unified approach to describing slip in gases and liquids in the range of moderate Knudsen numbers.

The structure of the paper is as follows. Section 2 discusses how the Navier boundary condition is introduced at the liquid-solid and liquid-liquid interfaces, and reveals the inadequacy of this approach when it comes to contact between two liquids.

In the 3rd Section, an analysis is made to reveal the duality of the pair of slip lengths at the liquid-liquid interface.

In the 4th Section, further development of the approach to slip boundary conditions allows to obtain new results on the motion of spherical drops of some special fluids.

In the 5th Section, it is shown that, in addition to the condition of tangential (transverse) partial slip, normal (longitudinal) slip also occurs at the interface between gas and liquid (or solid). Normal sliding is accompanied by a change in gas density. As follows from molecular kinetic theory, the slip length in a gas is of the order of the mean free path of a molecule in the gas. A formula is derived for the rise velocity of a small bubble, taking into account both normal and tangential slip at the interface.

In the 6th Section, an equation is derived for the terminal velocity of an aerosol ( $0.07 < \mathrm{Kn} < 0.3$ ), taking into account tangential and normal slipping in the air. The factor $A$ in the Cunningham correction is calculated. The result is compared with the experiment.

In the 7th Section, an experiment is proposed to determine the slip lengths in contacting liquids by measuring the velocity of a liquid droplet in another liquid.

## 2. Navier boundary conditions at the liquid-liquid interface

The Navier condition [11] is a mathematical formulation of the boundary conditions of slip-with-friction, in which the tangential component of the fluid flow velocity on a solid surface is proportional to the rate of strain or, equivalently, to the viscous stress tensor component that corresponds to this rate developing [7-9, 37]

The problem of the partial slip of a liquid droplet in an external fluid with the Navier boundary condition in a spherical coordinate system $(r, \theta, \varphi)$

$$|\Delta V_\theta| = Sl|\sigma_{r\theta}|, \tag{1}$$

which dates back to the work of Basset [12] is considered in [34]. Here $\Delta V_\theta$ is the tangential relative velocity of the droplet, $\sigma_{r\theta}$ is the tangential viscous shear stress on the spherical interface, and *Sl* is a 'slip coefficient' that can be determined empirically. Condition (1) is expanded as

$$V_\theta(R,\theta) - V_\theta'(R,\theta) = \frac{\beta R}{\eta}\sigma_{r\theta}(\theta, R) \quad , \tag{2}$$

where *R* is the radius of the liquid droplet; $V_\theta(R,\theta)$ and $V_\theta'(R,\theta)$ are the polar components of velocity of external and internal liquids, respectively;

$$\sigma_{r\theta}(\theta, R) = \eta\left(\frac{1}{r}\frac{\partial V_r}{\partial\theta} + \frac{\partial V_\theta}{\partial r} - \frac{V_\theta}{r}\right)_{r=R} , \tag{3}$$

where $\eta$ is the dynamical viscosity of the external liquid; $\beta$ is the dimensionless slip coefficient [34-36],

$$|\beta| = \frac{Sl}{R}\eta . \tag{4}$$

In Ref. [34], based on the results of Refs. [35], in the limit of small Reynolds numbers ( $\mathrm{Re} << 1$ ) for the force $F_0$ acting on a droplet moving in an external liquid with a constant velocity $V_0$, the expression was obtained (see Eqs. (75)-(76) in Ref. [34]):

$$F_0 = 2\pi R\eta V_0 \frac{2 + 3\kappa + 6\beta\kappa}{1 + \kappa + 3\beta\kappa}, \tag{5}$$

where $\kappa = \frac{\eta'}{\eta}$ is the ratio of the dynamic viscosity of the internal liquid (droplet), $\eta'$, to that of the external liquid, $\eta$. On the other hand, the force acting on a drop of liquid that rises or falls under the influence of gravity and buoyancy is given by

$$F_0 = \frac{4\pi}{3}(\rho' - \rho) g R^3, \tag{6}$$

where $\rho$ and $\rho'$ are the mass densities of external liquid and the liquid inside the droplet, respectively; $g$ is the acceleration of gravity. By equating Eqs. (5) and (6), one can obtain an expression

$$V_0 = \frac{2(\rho' - \rho) g R^2}{3\eta} \frac{1 + \kappa + 3\beta\kappa}{2 + 3\kappa + 6\beta\kappa}. \tag{7}$$

Eq. (7) describes the steady-state velocity of slow motion of a small spherical liquid droplet of radius $R$ in an external liquid. It is a solution of the Stokes equations

$$\eta \overrightarrow{\nabla^2} \mathbf{V} = \nabla p, \quad (\nabla \cdot \mathbf{V}) = 0, \tag{8}$$

$$\eta' \overrightarrow{\nabla^2} \mathbf{V}' = \nabla p', \quad (\nabla \cdot \mathbf{V}') = 0 \tag{9}$$

under the boundary condition (2). Here, the velocity $\mathbf{V}'$ and the pressure $p'$ describe the flow inside the liquid droplet.

Eq. (7) was also obtained in a different way in Ref. [36] (see Eq. (39) there) and the Supporting material to this paper, where this expression is obtained in a different way.

Note that under condition no-slip ($\beta$ = 0) Eq. (7) transforms into the Hadamard-Rybczynski equation (HRE), [18-19]:

$$V_{HR} = \frac{2(\rho' - \rho) g R^2}{3\eta} \frac{1 + \kappa}{2 + 3\kappa}, \tag{10}$$

HRE turns into the Stokes formula in the limit of infinite viscosity of the liquid drop ( $\eta' \to \infty$ , $\kappa \to \infty$ ), [20]:

$$V_S = \frac{2(\rho' - \rho) g R^2}{9\eta}. \tag{11}$$

Thus, a rigid body is treated as a liquid with infinite viscosity and, accordingly, an infinitely long relaxation time.

If $\kappa \to \infty$, i.e. the viscosity of the droplet is much greater than the viscosity of the surrounding liquid, and λ has a value limited in absolute value, Eq. (7) takes the form

$$V_B = \frac{2(\rho' - \rho) g R^2}{9\eta} \frac{1+3\beta}{1+2\beta}, \tag{12}$$

Eq. (12) describes the sliding of a solid sphere in an external fluid with viscosity η and slip parameter β. It coincides with the classical result of Basset [9, 12, 38]. At β=0, Eq. (12) becomes the Stokes formula (11).

If $\kappa \to 0$, i.e. the viscosity of the droplet is much smaller than the viscosity of the external liquid, HRE given by Eq. (10) takes the form

$$V_c = \frac{(\rho' - \rho) g R^2}{3\eta}. \tag{13}$$

Formula (13) is expected to describe the rise of a small air bubble ( $\rho' \to 0$ ) in a liquid [20].

There are two main approaches to interpreting the mechanism of momentum transfer at an interface in the presence of partial slip. In one of them, the value 1/β is considered from the same standpoint as the dimensionless coefficient of friction between solid contacting surfaces [12,34-35]. “This postulate is analogous to Coulomb's law of friction for solid surfaces moving relatively to each other", Ref. [35]. It is assumed that the friction coefficient should depend on the material properties of the liquids, and not on the characteristics of the velocity field.

Another approach, which goes back to the study of sliding of polymeric materials, introduces the concept of a thin intermediate layer near the interface with modified (reduced) viscosity, Refs. [39-41]. It is assumed that momentum transfer between the two phases occurs due to the viscosity of the intermediate layer. In this approach, instead of the slip coefficient $\beta$ , the so-called slip length is usually introduced, [7-9], which, following the Ref. [9], we will denote by $\lambda$.

When considering a liquid-solid interface with partial slip on the surface of a solid sphere (in this case, Eq. (1) is usually called the Navier condition), one must set $V_\theta'(R,\theta)=0$ in the boundary condition (2). At liquid-solid interface of arbitrary shape the tangential velocity $\mathbf{V}_\tau$ is determined by

$$\mathbf{V}_\tau = \frac{\lambda}{\eta}\left[\mathbf{I}-\mathbf{n}\otimes\mathbf{n}\right](\boldsymbol{\sigma}\cdot\mathbf{n}), \qquad (14)$$

$$(\mathbf{V}_\tau \cdot \mathbf{n}) = 0. \qquad (15)$$

where $\lambda$ is the slip length, $\boldsymbol{\sigma}$ is the viscous stress tensor, $\mathbf{n}$ denotes the normal to the interface, directed into the liquid. According to Eq. (14), the velocity of the liquid at the interface is determined by the component of the stress tensor corresponding to the shear rate in the corresponding direction. In the case of a flat sliding surface (Fig. 1a), this condition can be rewritten as

$$V_x(0) = \frac{\lambda}{\eta}\sigma_{xz}(0), \qquad (16)$$

where $\lambda$ is the slip length, $V_x(0)$ is the velocity of liquid at liquid-solid interface (z = 0).

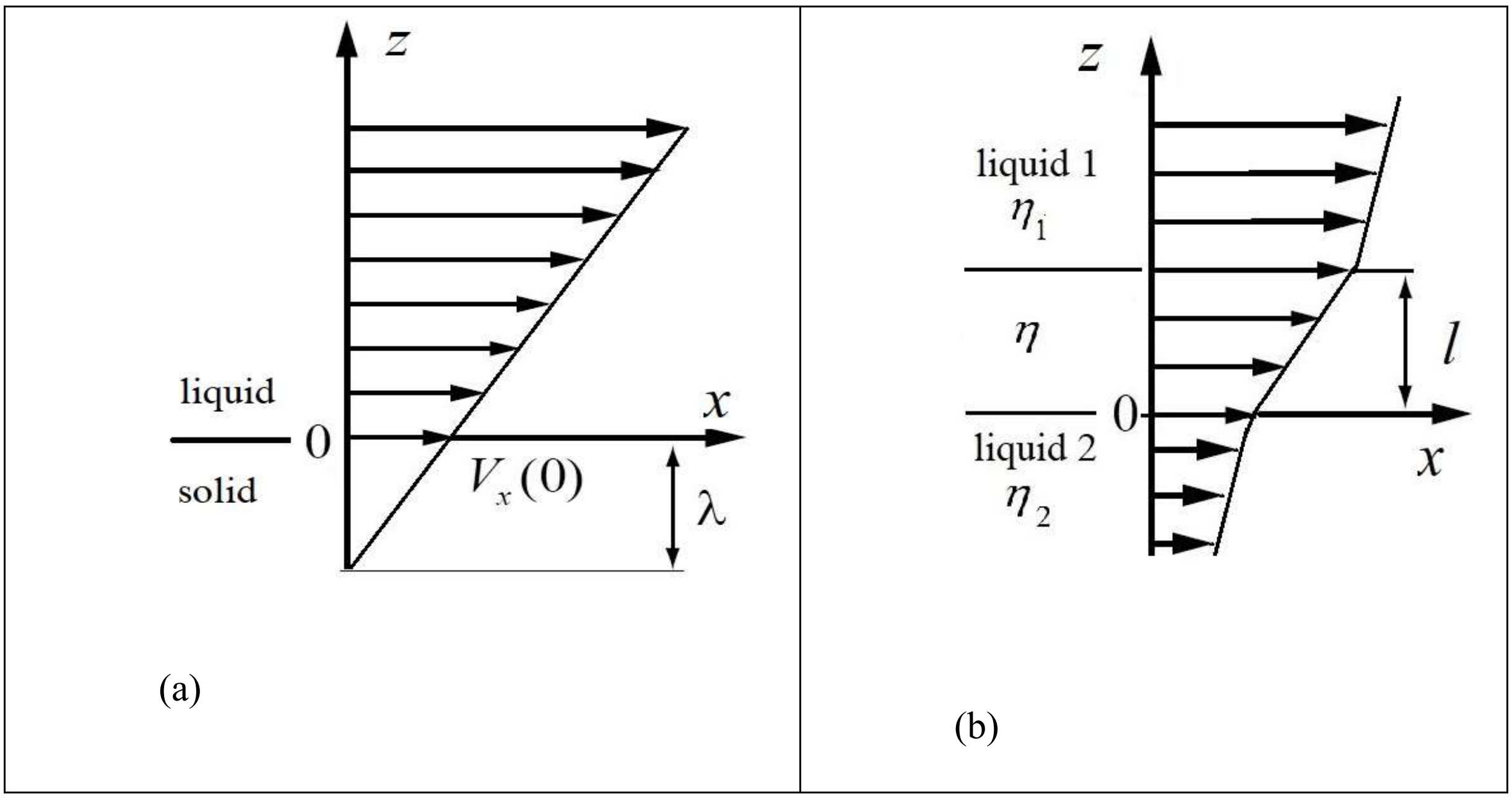

**Fig. 1**. Liquid velocity profile $V_x(z)$ near liquid-solid interface ($z$ = 0) with partial slip condition given by Eq. (18); λ is the slip length.

Considering that the tangential component of the viscous stress tensor has the form

$$\sigma_{xz} = \eta \frac{\partial V_x}{\partial z} , \tag{17}$$

then Eq.(14) can be rewritten as

$$V_x(0) = \lambda \left. \frac{\partial V_x(z)}{\partial z} \right|_{z=0} . \tag{18}$$

In the problems of fluid mechanics, the solid (rigid) state and the liquid differ only in viscosity. Indeed, let us assume that there is an interface of separation between two immiscible liquids, where one liquid slides over the other. If the viscosity of one of the liquids tends to infinity, there is a liquid-solid interface with a boundary condition of partial slipping. Thus, it is possible to transform a solid-liquid interface into a liquid-liquid interface by continuously changing the viscosity of one of the liquids. One can obtain a generalization of the Navier boundary condition to the liquid-liquid interface.

Indeed, molecular dynamics simulations of Couette and Poiseuille flows of two-layered liquid systems have shown that a partial sliding regime can exist, when there is a significant drop in average density in the liquid-liquid interface region [42]. In such case, the space immediately adjacent to the interface is proposed to be considered a region with altered (significantly lower than that of the surrounding liquids) interfacial viscosity $\eta(z)$, [39-40].

Then the velocity profile near the interface will have the form shown in Fig. 1b. Since the shear stress given by Eq. (6) must be constant in all three zones, one can write:

$$\eta_1 \frac{\partial V_{1x}}{\partial z}\bigg|_{z=l} = \eta(z)\frac{\partial V_x}{\partial z} = \eta_2 \frac{\partial V_{2x}}{\partial z}\bigg|_{z=0}, \tag{19}$$

where $V_{1x}$ and $V_{2x}$ are the velocities of the liquids 1 and 2, respectively. The velocity "jump" from $z=0$ to $z=l$ is

$$V_{1x}(l) - V_{2x}(0) = \eta(z)\frac{\partial V_x}{\partial z}\bigg|_{z=0} \int_0^l \frac{dz}{\eta(z)}. \tag{20}$$

As indicated in Ref. [11], on a macroscopic scale, the width of the transition zone between both bulk liquids is vanishingly small, $l \to 0$, and the viscosity $\eta$ of this diffuse region is also very small (although ratio, $l/\eta$, has a finite limit), so we can approximately move from Eq. (20) to the following expression:

$$V_{1x}(z \approx 0) - V_{2x}(0) = \alpha\sigma_{xz}(0) \tag{21}$$

or

$$V_{1x} - V_{2x} = \alpha\sigma_{xz}, \tag{22}$$

where α is slip parameter of the interface; the values of $V_{1x}$, $V_{2x}$, and $\sigma_{xz}$ are taken at the infinitely thin intermediate layer between two liquids, 1 and 2.

Comparing Eq.(11) and Eq.(5), we note that expression (22) can be rewritten similarly to condition (14) as follows [40-41]:

$$\mathbf{V}_{\tau 1} - \mathbf{V}_{\tau 2} = \alpha\left[\mathbf{I} - \mathbf{n}\otimes\mathbf{n}\right](\boldsymbol{\sigma}\cdot\mathbf{n}), \tag{23}$$

$$(\mathbf{V}_{\tau 1}\cdot\mathbf{n}) = (\mathbf{V}_{\tau 2}\cdot\mathbf{n}) = 0. \tag{24}$$

The left-hand side of Eq. (23) contains the relative tangential velocity at the liquid-liquid interface.

The above approaches to describing the partial sliding at a liquid-liquid interface require further development. First, it should be noted that the coefficient *Sl* in formula (1) can be nondimensionalized in two ways: through the viscosity of one or the other contacting liquid, so we have two dimensionless slip coefficients or two corresponding slip lengths. In a generalized approach, both liquids (and their properties) at the interface should be considered equally.

In the region of a liquid-liquid interface (Fig. 1 b), three zones can be generally distinguished: two liquids and a thin transition layer that can sometimes be interpreted by analogy with the dry friction of solid surfaces with all its attendant attributes, such as static friction, which depends on the compressive external pressure. Therefore, from a physical perspective, it would be appropriate to separate the friction coefficient of the surfaces (in a two-dimensional region) 1/β from the slip length, which describes the velocity profile in the three-dimensional volume of the adjoining liquids.

It is known that the partial slip condition is realized at the hydrophobic-hydrophilic solid-liquid interface, [8, 37, 43]. Similarly, systems that have a liquid-liquid interface where the partial slip condition is expected to be satisfied are water-oil and oil-water emulsions. There is an interface separating water and a hydrophobic liquid (oil).

## 3. Duality of the slip length at the liquid-liquid interface

Let us analyze the above-listed variants of partial slip boundary conditions at the liquid-liquid interface. First of all, we note that the case of a liquid-liquid interface differs from a liquid-solid one by the presence of two liquids, which means that there are two possibilities to nondimensionalize the *Sl* coefficient in Eq. (1) that gives two slip lengths that are different in magnitude but physically equal.

Indeed, condition (1), which relates the absolute value of the difference in the tangential velocities of two liquids separated by an interface to the modulus of the corresponding tangential component of the viscous stress tensor by means of a certain non-negative slip coefficient, is the most universal. However, when, eliminating symbols of modulus in Eq.(1), we move to the expression (2), a new requirement arises, related to the equivalence of the two liquids from the standpoint of physical description. The requirement of symmetry with respect to the replacement of one fluid by another requires the introduction of a second boundary condition, along with condition (2). Thus, in addition to Eq.(2), one can write a second boundary condition, where the quantities related to droplet are replaced by ones related to the exterior liquid:

$$V_\theta'(R,\theta) - V_\theta(R,\theta) = \beta' R\left(\frac{1}{r}\frac{\partial V_r'}{\partial\theta} + \frac{\partial V_\theta'}{\partial r} - \frac{V_\theta'}{r}\right)_{r=R}, \qquad (25)$$

where the second dimensionless slip coefficient β' is introduced.

Considering that the equality of the tangential components of the viscous stress tensor must be satisfied at the interface

$$\sigma_{r\theta} = \sigma_{r\theta}', \tag{26}$$

i.e.

$$\eta\left(\frac{1}{r}\frac{\partial V_r}{\partial \theta} + \frac{\partial V_\theta}{\partial r} - \frac{V_\theta}{r}\right) = \eta'\left(\frac{1}{r}\frac{\partial V_r'}{\partial \theta} + \frac{\partial V_\theta'}{\partial r} - \frac{V_\theta'}{r}\right), \tag{27}$$

and comparing Eq. (2) and Eq. (25) one can get the following relationship between two slip coefficients:

$$\frac{\eta}{\beta} = -\frac{\eta'}{\beta'} \quad \text{or} \quad \frac{\beta'}{\beta} = -\frac{\eta'}{\eta} = -\kappa. \tag{28}$$

Then, taking into account Eq. (4), we obtain

$$\beta' = -\frac{\eta' Sl}{R}. \tag{29}$$

Since $Sl \geq 0$ by definition, then $\beta' \leq 0$. Thus, at the liquid-liquid interface there are two slip coefficients, β and β'. One of them must be positive and the other negative.

Since in the problem under consideration there is not always a characteristic size *R*, for example, on a flat interface such a size is absent, it is convenient to introduce a pair of slip lengths, λ and λ', positive and negative, according to the definition:

$$\lambda = \beta R = \eta Sl, \qquad \lambda' = \beta' R = -\eta' Sl, \qquad \frac{\lambda'}{\lambda} = -\kappa. \tag{30}$$

Let there be two parallel rigid planes. The z-axis is directed upward perpendicular to the planes, so that one plane has coordinate $z = a$, the second $z = -b$, where $a>0$ and $b>0$. The upper plane is stationary, the lower one moves with a constant velocity $V$ (Fig.2).

Let us consider a plane flow (Couette flow) involving two immiscible fluids 1 (viscosity $\eta$) and 2 (viscosity $\eta'$), the interaction between which is carried out according to the partial slip condition Eq.(1) with a coefficient *Sl* at the interface at z=0. The pressure gradient, $\nabla p$, is zero.

$$\eta\overrightarrow{\nabla^2}\mathbf{V} = 0\,, \quad (\nabla\cdot\mathbf{V}) = 0\,, \tag{31}$$

$$\eta\frac{\partial^2 V_x(z)}{\partial z^2} = 0, \quad \frac{\partial V_x}{\partial x} = 0\,, \tag{32}$$

Similarly, for the liquid 2

$$\eta'\frac{\partial^2 V_x'(z)}{\partial z^2} = 0, \quad \frac{\partial V_x'}{\partial x} = 0\,, \tag{33}$$

The tangential component of the viscous stress tensor at the interface has the form (17). Therefore, the partial slip conditions for fluids 1 and 2 have the form Eq.(34) and Eq.(35), respectively:

$$\sigma_{xz} = \eta\left.\frac{\partial V_x}{\partial z}\right|_{z=0} = \frac{\eta}{\lambda}(V_x(0) - V_x'(0))\,, \tag{34}$$

$$\sigma'_{xz} = \eta'\left.\frac{\partial V_x'}{\partial z}\right|_{z=0} = \frac{\eta'}{\lambda'}(V_x'(0) - V_x(0))\,. \tag{35}$$

According to Eq. (26), viscous stresses Eq. (34) and Eq. (35) are identical, which yields relations (30).

The solution of Eqs. (32) and (33) for the flow velocity in fluids 1 and 2, taking into account the boundary conditions, has the form

$$V_x(z) = \frac{(\lambda - \kappa^{-1}a + z)V}{\lambda - b - \kappa^{-1}a}\,, \tag{36}$$

$$V_x'(z) = \frac{\kappa^{-1}(z-a)V}{\lambda - b - \kappa^{-1}a}\,. \tag{37}$$

Then, for the velocity difference at the interface, one can obtain

$$V_x(0) - V_x'(0) = \frac{\lambda V}{\lambda - b - \kappa^{-1}a}\,. \tag{38}$$

For a positive velocity $V$ in the coordinate axes shown in Fig. 2a, according to the concept of momentum transfer through viscous friction, the velocity of fluid 1 $V_x(z)$ should decrease with increasing z from –b to 0. In this case, the corresponding component of the viscous stress tensor (35) will be negative. Given equality (26), the derivative of the velocity of fluid 2 with respect to the z coordinate, $\frac{\partial V_x'}{\partial z}$, will also be negative. Therefore, the slope of the graphs describing the velocity profile will be exactly as shown in Fig. 2a,b.

However, the velocity difference at interface (38) can formally be either positive (Fig. 2a) or negative (Fig. 2b), or equal to zero if there is no slip at z=0. In the first case, (Fig. 2a), as follows from (34), we have $\lambda < 0$, and in the second case, (Fig. 2b), $\lambda > 0$.

We will verify this by analyzing formulae (36)-(38). Obviously, the velocity field shown in Fig. 2a corresponds to positive values of the velocity $V_x(z)$, as well as non-negative values of $V_x'(z)$ and their difference $V_x(0) - V_x'(0)$. Let us find which λ corresponds to the set of these conditions.

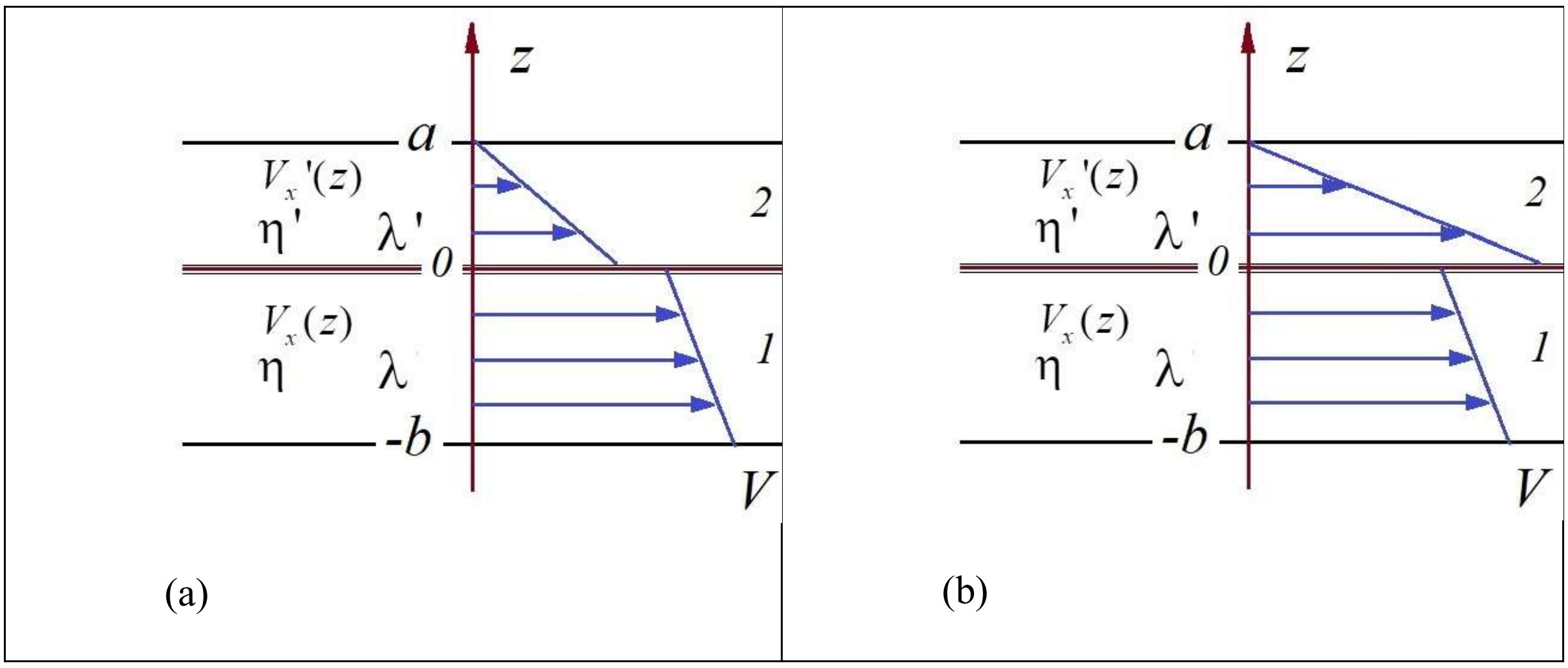

**Fig. 2.** Viscous flows of fluids 1 and 2 with partial slip condition at their interface, $z = 0$, initiated by motion of a plane $z = -b$ with velocity $V$ and the no-slip boundary condition at $z = -b$. The slip lengths in the liquids 1 and 2, respectively, are (a) $\lambda < 0$ and $\lambda' > 0$; (b) $\lambda > 0$ and $\lambda' < 0$.

It is easy to show that

$$V_x(z) > 0, \quad z \in [-b, 0], \text{ if } \lambda > \kappa^{-1}a + b \text{ or } \lambda < \kappa^{-1}a, \tag{39}$$

$$V_x'(z) \geq 0, \quad z \in [0, a], \text{ if } \lambda < b + \kappa^{-1}a. \tag{40}$$

$$V_x(0) - V_x'(0) \geq 0, \text{ if } \lambda > b + \kappa^{-1}a \text{ or } \lambda \leq 0. \tag{41}$$

The combination of conditions (39)-(41) results in

$$\lambda \leq 0. \tag{42}$$

On the other hand, if, when conditions (39)-(40) are satisfied, the velocity difference (38) is less than zero, which corresponds to Fig. 2b, then condition (41) should be replaced by:

$$V_x(0) - V_x'(0) < 0, \text{ if } \lambda < b + \kappa^{-1}a \text{ and } \lambda > 0. \tag{43}$$

The combination of conditions (39)-(41) and (43) results in

$$\lambda \in [0, \kappa^{-1}a], \text{ i.e., } \lambda > 0. \tag{44}$$

It can be argued that steady-state flow with partial slip at the interface, satisfying condition (44) or Fig. 2b, is not realized in practice. To explain this, let us consider how the energy supplied to the system under consideration is dissipated.

Let us denote the external shear force acting on a unit area of the lower plane by $\sigma$. Then, the power supplied to a unit area is equal to $\sigma V$ (Fig. 2). This power is released as heat in the bulk of the fluids due to viscosity and at the interface between the two fluids due to friction.

The dissipation of kinetic energy in an incompressible fluid is determined by [44]

$$\dot{E}=-\frac{\eta}{2}\int\left(\frac{\partial V_i}{\partial x_k}+\frac{\partial V_k}{\partial x_i}\right)^2 dV=-\frac{1}{2\eta}\int\sigma_{ik}\sigma_{ik}dV, \quad (45)$$

which in the case under consideration, taking into account $\sigma_{ik}=\sigma_{ki}$, gives the absorbed power per unit length of the system along the axis of motion

$$N_V=\eta\int_{-b}^{0}\left(\frac{\partial V_x(z)}{\partial z}\right)^2 dz+\eta'\int_{0}^{a}\left(\frac{\partial V_x'(z)}{\partial z}\right)^2 dz. \quad (46)$$

Differentiating Eq. (36) and Eq. (37) with respect to *z*, we obtain

$$\frac{\partial V_x}{\partial z}=\frac{V}{\lambda-b-\kappa^{-1}a}, \qquad \frac{\partial V_x'}{\partial z}=\frac{\kappa^{-1}V}{\lambda-b-\kappa^{-1}a}. \quad (47)$$

Then

$$N_V=\frac{(b+\kappa^{-1}a)\eta V^2}{(b+\kappa^{-1}a-\lambda)^2}. \quad (48)$$

The power released per unit area of the interface due to friction, which is equal to the scalar product of the shear force and the velocity of displacement of one surface relative to the other. The shear force is directed along the x-axis, as is the velocity of displacement of the bottom plate *V*. The power dissipated by friction at the interface is equal to

$$N_S=\sigma(V_x(0)-V_x'(0)). \quad (49)$$

The external force in the bulk fluid is compensated by the corresponding component of viscous stress. According to Newton's first law and taking into account Eq. (34) and Eq. (38), we have

$$\sigma=-\sigma_{xz}=-\frac{\eta}{\lambda}(V_x(0)-V_x'(0))=-\frac{\eta V}{\lambda-b-\kappa^{-1}a}. \quad (50)$$

Substituting Eq. (50) into Eq. (49) and taking into account Eq. (38), we have

$$N_S = -\frac{\eta\lambda V^2}{(\lambda - b - \kappa^{-1}a)^2}. \quad (51)$$

The total energy dissipation is the sum of the volumetric and surface contributions:

$$N = N_V + N_S = \frac{\eta V^2}{b + \kappa^{-1}a - \lambda}. \quad (52)$$

Taking into account Eq. (50), this result can be represented as the product of the external shear force applied to the lower plane and the velocity of this plane *V*, which obviously expresses the power supplied by the external force:

$$N = \sigma V. \quad (53)$$

This result is independent of $\lambda$ and therefore holds for both Fig. 2(a) and Fig. 2(b). The criterion for selecting the correct velocity profile is the sign of the power dissipated at the interface. Since the friction process is accompanied by heat release, then $N_S \geq 0$, which, taking into account Eq. (51), implies $\lambda \leq 0$. Therefore, the correct velocity distribution corresponds to Fig. 2(a). The velocity distribution shown in Fig. 2b, for which $N_S < 0$, is not realized in stationary processes.

Thus

$$\lambda = -Sl\eta. \quad (54)$$

Negative slip length (54) characterizes the fluid 1. Then fluid 2, in accordance with Eq. (30), will have a positive-definite slip length:

$$\lambda' = -\kappa\lambda = Sl\eta'. \quad (55)$$

Note also that Eqs. (36)-(37) can be rewritten in equivalent form concluding $\lambda'$, using Eq. (55), which reflects the symmetry of the partial slip boundary condition with respect to the fluids 1 and 2:

$$V_x(z) = -\frac{(\kappa z - a - \lambda')V}{\lambda' + \kappa b + a}, \qquad V_x'(z) = \frac{(a-z)V}{\lambda' + \kappa b + a} \; . \tag{56}$$

In the example discussed, we saw that fluid 1, directly connected to the friction source, has a negative sliding length, while the driven fluid has a positive one. However, if we choose the opposite direction for the z-axis in Fig. 2, the signs of the derivatives $\frac{\partial V_x}{\partial z}$ and $\frac{\partial V_x}{\partial z}$ will be reversed. Therefore, for the case depicted in Fig. 2(a), we should assume $\lambda > 0$ and $\lambda' < 0$, while for Fig. 2(b), we have $\lambda < 0$ and $\lambda' > 0$. Therefore, in the example considered, the choice of the signs of the sliding lengths is tied to the choice of the coordinate system and is relative. Of course, this cannot change the physically meaningful conclusion that, in reality, system given by Fig. 2(a) is realized.

To summarize, based on thermodynamics, we can formulate the following rule: at a liquid-liquid interface with partial sliding, the liquid transmitting the impulse of the external force to the second liquid through the interface has a leading velocity in the direction of motion; This, taking into account the choice of coordinate system, distributes the signs between the sliding lengths of the contacting liquids in the formula $\lambda' = -\kappa\lambda$, and for the ratio of viscosities the inequality $\kappa > 0$ is always satisfied.

## 4. Spherical Droplet sliding

Let us determine the sign distribution in the pair of slip lengths between a liquid droplet and an external fluid. The correct choice of signs should ensure positive energy dissipation due to friction at the interface between the two fluids.

For the sake of clarity, let us assume that the droplet rises. Consider it in a coordinate system in which it is at rest as a whole [Fig. 3(a)]. The external liquid moves downward, and the boundary condition at infinity for the external fluid is determined by the droplet rise velocity, taken with the opposite sign. This boundary condition at infinity determines the boundary condition at the droplet surface, taking into account the partial slip effect.

According to the rule established in Section 2, the leading fluid must have a higher velocity at the interface than the trailing fluid. In the system under consideration, the leading fluid is the external fluid, which is directly related to the boundary conditions at infinity, while the droplet's fluid is the trailing fluid. Therefore, the correct relationship between the velocities of the two fluids in contact at the interface is shown in Fig. 3(a) in the droplet resting frame. Fig. 3(b) depicts the same situation in a system in which the vessel containing the external fluid is stationary (assuming it has very large or infinite dimensions).

The resulting image allows us to establish the relationship between the signs of the slip lengths in the internal and external fluids.

The external force acting on a unit area of the droplet surface from the external liquid is related to the corresponding component of the viscous stress tensor by formula (25), which can be rewritten by analogy with Eq. (34) in terms of the slip length.

$$\sigma_{r\theta}(R,\theta) = \frac{\eta}{\lambda}(V_\theta(R,\theta) - V_\theta{}'(R,\theta)) . \tag{57}$$

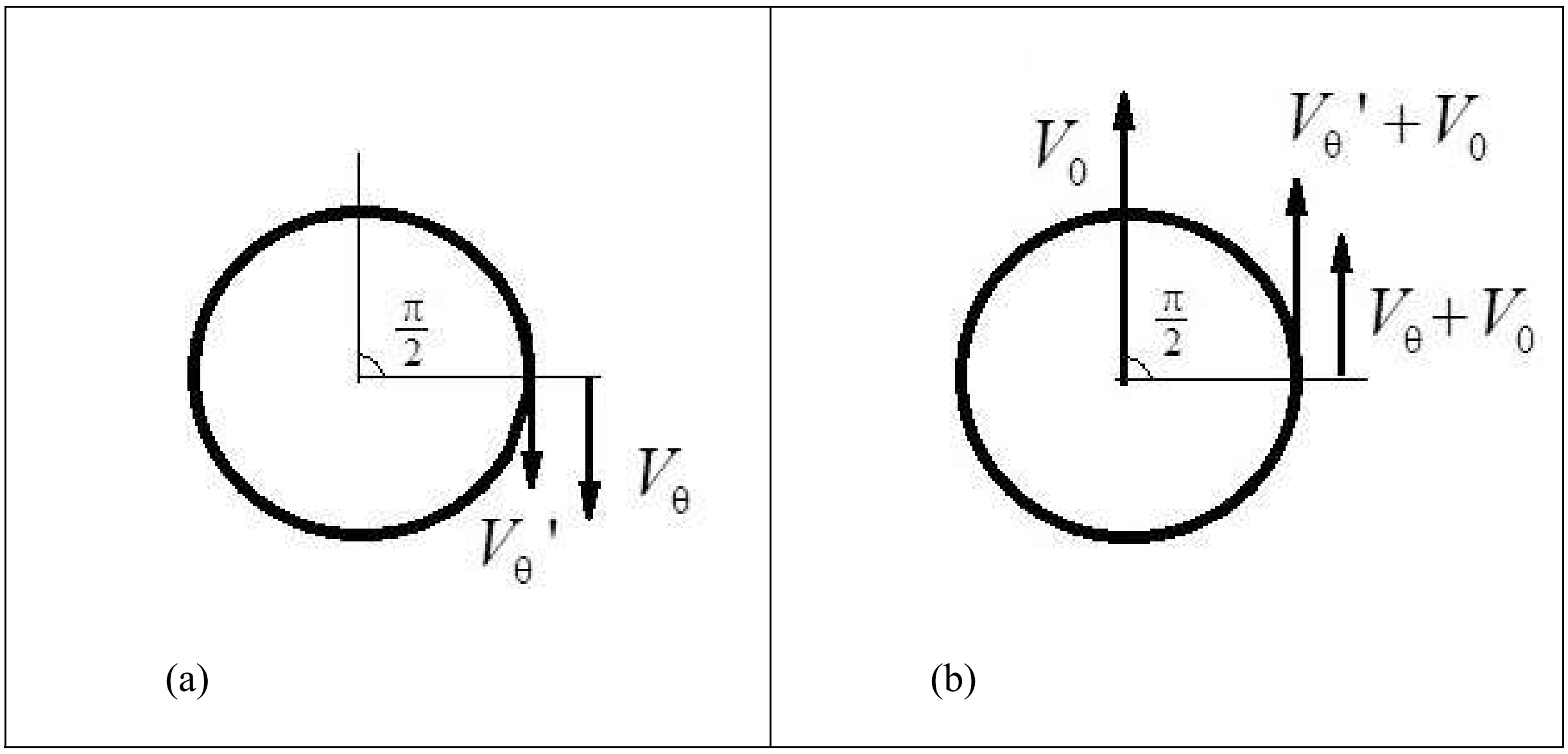

**Fig. 3**. Direction of velocities on the droplet surface in the presence of slip at the interface in the coordinate system in which the droplet is at rest (a), and in which it rises with velocity $V_0$ (b).

This force is applied to the droplet's surface. It counteracts the buoyancy force that causes the droplet to rise.

As follows from Fig. 2(a), the algebraic value of the shear rate caused by the surface force (57) should be taken as $V_\theta(R,\theta) - V_\theta'(R,\theta)$. Therefore, the power released per unit area of the interface is determined by the formula

$$N_S = \sigma_{r\theta}(R,\theta)(V_\theta(R,\theta) - V_\theta'(R,\theta)) = \frac{\eta}{\lambda}(V_\theta(R,\theta) - V_\theta'(R,\theta))^2 . \tag{58}$$

For (58) to be positive, the slip length in the external liquid must be positive; in this case, the slip length of the liquid in the droplet will be negative:

$$\lambda > 0, \quad \lambda' = -\kappa\lambda < 0 . \tag{59}$$

It can be shown that, in this formulation of the problem, the rise velocity of a liquid droplet, $V_0$, is determined by formula (S.66) from Supporting Material, which has the form:

$$V_0 = \frac{(\rho' - \rho) g R^2}{3\eta} \left\{ 1 - \left( \frac{6\lambda}{R} + 3 + \frac{2\eta}{\eta'} \right)^{-1} \right\} \tag{60}$$

where λ is the slip length of the external liquid. The liquid flow is described by the following equations [see Eqs. (S70)-(S77), Supporting material]:

1) The external liquid ($r > R$)

$$V_r = \frac{(\rho' - \rho) g R^2}{3\eta} \left\{ 1 - \frac{R}{r} - \left( \frac{6\lambda}{R} + 3 + \frac{2\eta}{\eta'} \right)^{-1} \left[ 1 - \left( \frac{R}{r} \right)^3 \right] \right\} \cos\theta, \tag{61}$$

$$V_\theta = \frac{(\rho' - \rho) g R^2}{6\eta} \left\{ \frac{R}{r} - 2 + \left( \frac{6\lambda}{R} + 3 + \frac{2\eta}{\eta'} \right)^{-1} \left[ 2 + \left( \frac{R}{r} \right)^3 \right] \right\} \sin\theta, \tag{62}$$

$$p = -\frac{(\rho' - \rho) g R^3}{3 r^2} \cos\theta. \tag{63}$$

Stream function is

$$\Psi(r, \theta) = \frac{(\rho' - \rho) g R^4}{6\eta} \sin^2\theta \left\{ \left( \frac{r}{R} \right)^2 - \frac{r}{R} + \left( \frac{6\lambda}{R} + 3 + \frac{2\eta}{\eta'} \right)^{-1} \left( \frac{R}{r} - \left( \frac{r}{R} \right)^2 \right) \right\}. \tag{64}$$

2) The internal liquid (r < R)

$$V_r' = -\frac{(\rho' - \rho) g R^2}{3\eta'} \left( \frac{6\lambda}{R} + 3 + \frac{2\eta}{\eta'} \right)^{-1} \left\{ 1 - \left( \frac{r}{R} \right)^2 \right\} \cos\theta, \tag{65}$$

$$V_\theta' = -\frac{(\rho' - \rho) g R^2}{3\eta'} \left( \frac{6\lambda}{R} + 3 + \frac{2\eta}{\eta'} \right)^{-1} \left\{ 2 \left( \frac{r}{R} \right)^2 - 1 \right\} \sin\theta, \tag{66}$$

$$p' = \frac{10 (\rho' - \rho) g r}{3} \left( \frac{6\lambda}{R} + 3 + \frac{2\eta}{\eta'} \right)^{-1} \cos\theta. \tag{67}$$

Stream function is:

$$\Psi' = \frac{(\rho' - \rho) g R^4}{6\eta'} \sin^2\theta \left( \frac{6\lambda}{R} + 3 + \frac{2\eta}{\eta'} \right)^{-1} \left\{ \left( \frac{r}{R} \right)^4 - \left( \frac{r}{R} \right)^2 \right\}. \tag{68}$$

Taking into account Eq. (30), namely, $\lambda = -\kappa^{-1}\lambda'$, Eq. (60) can be rewritten in terms of the slip length of the liquid filling the drop, λ':

$$V_0 = \frac{(\rho' - \rho) g R^2}{3\eta} \left\{ 1 - \left( 3 + \frac{2\eta}{\eta'} \left[ 1 - \frac{3\lambda'}{R} \right] \right)^{-1} \right\}. \tag{69}$$

Eq. (60) can be also transformed, taking into account Eqs. (28):

$$V_0 = \frac{2(\rho' - \rho) g R^2}{3\eta} \frac{1 + \kappa + 3\kappa\lambda R^{-1}}{2 + 3\kappa + 6\kappa\lambda R^{-1}}, \qquad \kappa = \frac{\eta'}{\eta}, \tag{70}$$

or

$$V_0 = \frac{2(\rho' - \rho) g R^2}{3\eta} \frac{1 + \kappa - 3\lambda' R^{-1}}{2 + 3\kappa - 6\lambda' R^{-1}}. \tag{71}$$

By substituting $\lambda = \beta R$ [see Eqs.(28)], Eq. (70) transforms to Eq.(7).

Note that under the conditions of the problem under consideration, the Reynolds number satisfies the following relations

$$\mathrm{Re} = \frac{2\rho R V_0}{\eta} << 1. \tag{72}$$

Formula (7) and its equivalent Eqs. (60), (70), (71) can be called generalized Hadamard-Rybczynski equations.

The HRE, Eq. (10), is usually applied to describe the motion of a small droplet of liquid in another liquid. In its derivation, the no-slip boundary condition at the liquid-liquid interface was used [20]. This somewhat contradicts the initial assumption that both liquids are immiscible (poorly soluble in each other). Therefore, a more natural and generalizing condition is the partial slip of one liquid on the surface of the other one given by Eqs. (1)-(4), which introduce a modification of the Navier condition originally proposed for the liquid-solid interface. A generalized HRE given by Eq.(7) can be transformed into the usual HRE (10) at λ=0.

The conditions of applicability of the generalization of the HRE are the same as for the no-slip model. First of all, this is the condition $\mathrm{Re} \ll 1$ under which the linearized Navier-Stokes equations are valid. For raising or falling droplets, the velocity of which is determined by Eq. (70), this condition limits the maximum size of the droplet and, taking into account Eq. (72), has the form

$$R^3 \ll \frac{\eta^2}{\rho|\rho-\rho'|g}\frac{2\alpha+3+6\lambda R^{-1}}{\alpha+1+3\lambda R^{-1}}. \tag{73}$$

Reducing the inequality by extracting the cubic root, we obtain

$$R < \left(\frac{\eta^2}{\rho|\rho-\rho'|g}\frac{2\alpha+3+6\lambda R^{-1}}{\alpha+1+3\lambda R^{-1}}\right)^{\frac{1}{3}} \tag{74}$$

Given that the second fraction in parentheses in Eq. (74), as a rule, has a value of the order of 1, the expression can be used to roughly estimate the drop radius

$$R < \left(\frac{\eta^2}{\rho|\rho-\rho'|g}\right)^{\frac{1}{3}}. \tag{75}$$

The equation (70), which can be rewritten as

$$V_0 = \frac{2(\rho'-\rho)gR^2}{3\eta}\frac{\eta+\eta'+3\lambda\eta' R^{-1}}{2\eta+3\eta'+6\lambda\eta' R^{-1}}. \tag{76}$$

of course, does not exhaust all possible variants of boundary conditions, and therefore cannot claim universality, but there is a fairly wide and important class of substances for the description of which this equation can be applied. From a mathematical point of view, it is obvious that this is approximately the range of substances for the description of which the Boussinesq equation can be used [20, 45]:

$$V_0 = \frac{2(\rho'-\rho)gR^2}{3\eta}\frac{\eta+\eta'+2e(3R)^{-1}}{2\eta+3\eta'+2eR^{-1}}, \tag{77}$$

where $e$ is the "coefficient of surface viscosity" introduced by Boussinesq. Indeed, the structure of expression (77) almost coincides with the expression (76). In this case, the concept of the slip length seems to make more physical sense.

At typical values of parameters $\eta \sim 10^{-3}\,\mathrm{Pa\cdot s}$, $\rho \sim 10^3\,\mathrm{kg\cdot m^{-3}}$, $|\rho - \rho'| \sim 100\,\mathrm{kg\cdot m^{-3}}$, we have a limitation $R < 10^{-4}\,\mathrm{m}$, i.e., it is an emulsion containing very small droplets with radius of 100 microns or less. These can be, for example, oil-in-water or water-in-oil emulsions. In such systems with a hydrophobic-hydrophilic liquid-liquid interface, partial slip of liquid particles should be expected in the dispersed phase. The study of electrophoresis or sedimentation processes in such emulsions can provide information about the value of slip length $\lambda$.

If highly viscous liquids with viscosity $\eta \sim 1\,\mathrm{Pa\cdot s}$, such as glycerin or castor oil, are used as the external liquid, then according to Eq. (75), within the framework of this concept, it is permissible to consider the motion of drops several mm in size.

A different asymptotic behavior occurs in the obtained formula (76) in the limit $\lambda \to \infty$: in this case, Eq. (76) takes the form given by Eq. (13). An infinitely long slip length is the absolute slip mode at the liquid-liquid interface. The HRE (10) has the same limit, but at $\eta \to \infty$, i.e., when the viscosity of the medium is many times greater than the viscosity of the droplet moving in it. This is physically a completely different condition. Note that Eq. (76) takes on the same form for finite slip length $\lambda$ at $\eta \to \infty$, as it should.

Let us consider the case where the slip coefficient in condition (1) is very large:

$$Sl \to \infty. \tag{78}$$

The slip coefficient *Sl*, according to Eqs. (30) and (45), is related to the slip lengths, taking into account the sign distribution found above, by the relations

$$Sl = \frac{\lambda}{\eta} = -\frac{\lambda'}{\eta'} . \tag{79}$$

Eq. (79) can be rewritten as

$$\lambda\eta' = -\eta\lambda' = \gamma \geq 0 , \tag{80}$$

where γ is a constant characterizing the interface, with dimensions $\mathrm{kg \cdot s^{-1}}$.

Using Eq. (80), one can rewrite Eq. (79) as

$$Sl = \frac{\gamma}{\eta\eta'} . \tag{81}$$

Given Eq. (81), condition (78) will be satisfied in two practically important cases:

1) if $\gamma \to \infty$, and the product $\eta\eta'$ is limited (the case of two liquids);

2) if γ and η are finite quantities, and $\eta' \to 0$ (the case when considering the rise of a gas bubble). The constant γ can be bounded rather than infinite if it is a finite product of a very large slip length λ and a very small viscosity coefficient η' (the droplet resembles a gas bubble) or, conversely, a very small λ and a very large η' (when the droplet resembles a solid).

Substituting Eq. (80) into Eq. (57) yields the following expression for the droplet velocity.

$$V_0 = \frac{(\rho' - \rho)gR^2}{3\eta}\left\{1 - \left(\frac{6\gamma + 2R\eta}{R\eta'} + 3\right)^{-1}\right\} . \tag{82}$$

Considering the above options, for normal liquids in the limit (78), Eq. (82) becomes formula (13), i.e., the motion of a droplet of the same density in a medium of the same viscosity occurs 1.5 times faster than the motion of a solid ball of the same size and density (11). Eqs (58)-(65) take the form:

1) The external liquid

$$V_r = V_c\left\{1-\frac{R}{r}\right\}\cos\theta, \tag{83}$$

$$V_\theta = V_c\left\{\frac{R}{2r}-1\right\}\sin\theta, \tag{84}$$

$$p = -\frac{\eta}{R}V_c\left(\frac{R}{r}\right)^2\cos\theta, \tag{85}$$

$$\Psi(r,\theta) = \frac{V_c R^2}{2}\sin^2\theta\left\{\left(\frac{r}{R}\right)^2-\frac{r}{R}\right\}. \tag{86}$$

where $V_c$ is determined by Eq. (13).

2) The internal liquid

$$V_r' = V_\theta' = 0, \qquad p' = 0. \tag{87}$$

It can be shown that the same solution, i.e., Eqs. (13), (83)-(87), is obtained by choosing the condition of complete slip at the interface in the form:

$$\sigma_{r\theta}(\theta,R) = \sigma_{r\theta}'(\theta,R) = 0. \tag{88}$$

Let us consider the directions of the velocities of the external and internal fluids on the droplet surface ($r = R$). Using Eq. (62) and Eq. (66), we find

$$V_\theta(R,\theta) - V_\theta'(R,\theta) = \lambda\frac{(\rho-\rho')gR}{\eta}\left(\frac{6\lambda}{R}+3+\frac{2\eta}{\eta'}\right)^{-1}\sin\theta. \tag{89}$$

For $\rho-\rho'>0$ и $\lambda>0$ the difference in tangential velocities given by Eq.(89), given that the angle $\theta$ is measured clockwise, is positive (Fig. 3).

For a sliding solid sphere, one can obtain

$$V_\theta(R,\theta) - V_\theta'(R,\theta) = \lambda\frac{(\rho-\rho')gR}{3\eta}\left(\frac{2\lambda}{R}+1\right)^{-1}\sin\theta. \tag{90}$$

For $\lambda = 0$ Eq. (89) takes the form of a no-slip boundary condition, $V_\theta(R,\theta) - V_\theta'(R,\theta) = 0$, as expected.

**5. Tangential and normal partial slip at the gas-liquid interface**

Let us prove that for a gas, unlike an incompressible liquid, a normal (longitudinal) slip condition have to be introduced at the interface with the dense phase (liquid or solid) in addition to tangential slip. Consider the gas flow density, which, generally speaking, consists of convective and diffusion components:

$$\mathbf{j} = n\mathbf{V}' - D\nabla n\,, \tag{91}$$

where $D$ is the self-diffusion coefficient of the gas molecules in the bubble, $\mathbf{V}'$ is the gas-dynamic velocity responsible for gas convection, and $n$ is the volume concentration of gas molecules

Let us represent the gas concentration in the bubble as follows:

$$n(r,\theta) = n_0 + \Delta n(r,\theta), \tag{92}$$

where $n_0$ is the volume-averaged concentration of the gas molecules. Then

$$\mathbf{j} = n_0\mathbf{V}' + \mathbf{V}'\Delta n - D\nabla\Delta n\,. \tag{93}$$

The steady-state flow satisfies the continuity relation:

$$(\nabla\cdot\mathbf{j}) = n\nabla\cdot\mathbf{V}' + \mathbf{V}'\nabla\Delta n - D\nabla^2\Delta n = 0\,. \tag{94}$$

Assume, that $|\Delta n(r,\theta)| << n_0$, $|\nabla\Delta n| \propto \frac{|\Delta n|}{R}$, $|\mathbf{V}'| \propto V_0$, where $V_0$ is the characteristic velocity, conveniently chosen to be the bubble's rise velocity.

Let us suppose that the incompressible fluid approximation is valid for convective gas motion

$$\nabla \cdot \mathbf{V}' = 0. \qquad (95)$$

Substituting Eq. (95) into Eq. (94), one can obtain the well-known diffusion equation with convection, [20]:

$$\mathbf{V}'\nabla\Delta n - D\nabla^2\Delta n = 0. \qquad (96)$$

We emphasize that condition (95) is necessary for Eq. (96) to be true.

Let us assume that for a given bubble size, diffusion dominates over convection, i.e., the inequality holds for the Peclet number.

$$\mathrm{Pe} = \frac{V_0 R}{D} << 1. \qquad (97)$$

Then Eq. (96) takes the form of a stationary diffusion equation

$$\nabla^2\Delta n \approx 0, \qquad (98)$$

and Eq. (91) can be rewritten as

$$\mathbf{j} \approx n_0\mathbf{V}' - D\nabla\Delta n. \qquad (99)$$

Consider a two-dimensional gas flow described by a gas-dynamic velocity $\mathbf{V}'$ (Fig. 4) near a flat liquid wall. Fig. 4 shows a plane in the gas, parallel to the wall, located at a distance equal to the mean free path of a molecule in the gas, $L$ The plane has a coordinate of 0 on the $z$-axis.

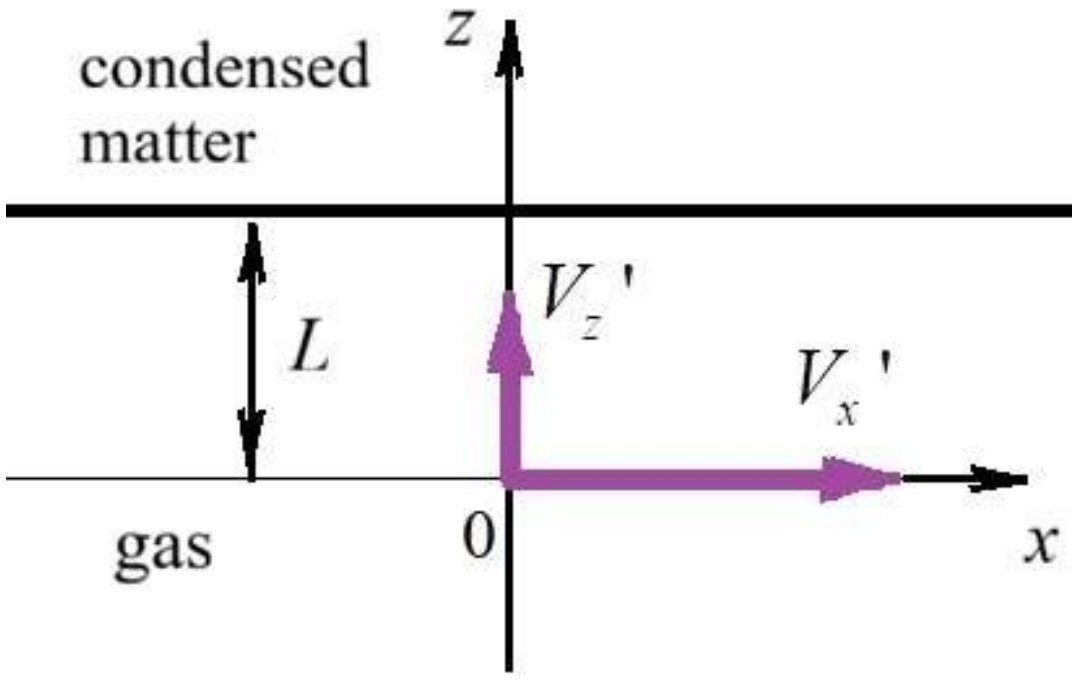

**Fig. 4**. Two-dimensional gas flow $(V_x', V_z')$ near a flat wall; $L$ is the mean free path of gas molecules.

Let us find the force c acting on the gas due to its interaction with the wall (a stationary plane of condensed matter). Consider a molecule whose coordinates $(x,0)$ at a given moment in time belong to this plane. The velocity of the molecule is composed of its thermal velocity ***v***, which has the Maxwellian statistical distribution corresponding to a given temperature *T*, and the gas-dynamic velocity $\mathbf{V}'(x,0)$:

$$\mathbf{U}(x,0) = \mathbf{v} + \mathbf{V}'(x,0)\,. \tag{100}$$

The average value of the thermal velocity modulus is many is orders of magnitude greater than the value of the gas-dynamic velocity.

We will consider the projection of velocity onto the z-axis. To simplify the notation, we will omit the coordinates $(x,0)$ in the equations from this section, assuming that we are talking about the gas-dynamic velocity and viscous stress in the gas immediately near the wall. The concentration of molecules whose thermal velocity is in the interval $[v_z, v_z + dv_z]$ is determined by the Maxwell distribution (we assume the gas is near thermodynamic equilibrium):

$$dn = n\left(\frac{m}{2\pi k_B T}\right)^{\frac{1}{2}} \exp\left(-\frac{m v_z^{\,2}}{2k_B T}\right) dv_z\,. \tag{101}$$

where $m$ is the mass of the gas molecule, and $k_B$ is the Boltzmann constant.

A gas molecule, upon colliding with a stationary liquid wall, transfers momentum $mU_z$ to it, where

$$U_z = v_z + V_z\,', \tag{102}$$

and the molecule itself loses this momentum. The molecule then leaves the liquid, having only thermal velocity, so the impact with the liquid can be classified as inelastic. Thus, we have two stages of interaction: collision and reflection.

The number of collisions per unit area per unit time is equal to $U_z dn$. Then the momentum transferred by the liquid wall to the gas molecules, having a velocity projection perpendicular to the interface, per unit area per unit time when falling on the wall, is determined by the expression

$$d\sigma_{1zz} = -mU_z^{\,2} dn = -mn\left(\frac{m}{2\pi kT}\right)^{\frac{1}{2}} U_z^{\,2} \exp\left(-\frac{m v_z^{\,2}}{2kT}\right) dv_z\,. \tag{103}$$

Integrating Eq. (103) taking into account Eq. (102), we obtain the transferred momentum per unit area per unit time:

$$\sigma_{1zz} = -mn\left(\frac{m}{2\pi k_B T}\right)^{\frac{1}{2}} \int_0^{\infty} \left(v_z + V_z\,'\right)^2 \exp\left(-\frac{m v_z^{\,2}}{2kT}\right) dv_z\,. \tag{104}$$

Assuming that the term proportional to $V_z\,'^2$ can be neglected, since the average thermal velocity given by [46]

$$\widehat{v_z} = \left( \frac{8k_B T}{\pi m} \right)^{\frac{1}{2}} \tag{105}$$

is much greater than the gas-dynamic velocity $V_z'$, we have

$$\sigma_{1zz} = -\frac{1}{2} n k_B T - mn \left( \frac{m}{2\pi k_B T} \right)^{\frac{1}{2}} \frac{2k_B T}{m} V_z' = -\frac{1}{2} n k_B T - \frac{3\eta'}{2L} V_z', \tag{106}$$

where, in accordance with molecular kinetic theory, the dynamic viscosity of the gas has the form

$$\eta' = \frac{mn}{3} \widehat{v_z} L, \tag{107}$$

where $mn = \rho'$ is the mass density of the gas.

Then the molecule detaches from the wall back into the gas phase, having a velocity distribution corresponding to a stationary wall. In this case, instead of Eq. (103), we have $d\sigma_{2zz} = -m v_z^2 dn$. Similar to Eq. (104), one can obtain

$$\sigma_{2zz} = -\frac{1}{2} n k_B T. \tag{108}$$

Adding Eq. (106) and Eq. (108), we obtain the momentum transferred to the gas by the wall per unit time per unit area during the inelastic collisions and thermalized reflections of gas molecules.

$$\sigma_{zz} = \sigma_{1zz} + \sigma_{2zz} = -n k_B T - \frac{3\eta'}{2L} V_z'. \tag{109}$$

Considering that the normal component of the viscous stress tensor is determined by the formula [44]

$$\sigma_{zz}' = -p_0' - p' + 2\eta' \frac{\partial V_z'}{\partial z}, \tag{110}$$

where $p_0{}'$ and $p'$ are the static and dynamic contributions to the pressure, respectively. Then, comparing Eq. (110) with Eq. (109), we obtain

$$-p_0{}'-p'+2\eta'\frac{\partial V_z{}'}{\partial z}=-nk_BT-\frac{3\eta'}{2L}V_z{}'. \quad (111)$$

Similarly, considering the inelastic transfer of tangential momentum corresponding to the velocity $V_x'(x,0)$, one can obtain

$$d\sigma_{xz}{}'=-mU_zV_x{}'dn=-mn\left(\frac{m}{2\pi kT}\right)^{\frac{1}{2}}U_zV_x{}'\exp\left(-\frac{mv_z{}^2}{2kT}\right)dv_z. \quad (112)$$

Then

$$\sigma_{xz}{}'\approx-mn\left(\frac{m}{2\pi k_BT}\right)^{\frac{1}{2}}V_x{}'\int_0^\infty v_z\exp\left(-\frac{mv_z{}^2}{2kT}\right)dv_z=-\frac{3\eta'}{4L}V_x{}'. \quad (113)$$

On the other hand, the viscous tangential viscous stress tensor component is determined by

$$\sigma_{xz}{}'=\eta'\left(\frac{\partial V_x{}'}{\partial z}+\frac{\partial V_z{}'}{\partial x}\right), \quad (114)$$

where the second term in the brackets in the system under consideration is equal to zero. Therefore, comparing Eq. (113) and Eq. (114), one can obtain

$$\frac{\partial V_x{}'}{\partial z}=-\frac{3V_x{}'}{4L}. \quad (115)$$

Taking into account Eq. (35), it follows from Eq. (115)

$$\lambda'=-\frac{4}{3}L. \quad (116)$$

Note that if we were to similarly consider the gas on the other side of the plane shown in Fig. 4, so that the molecules would fall onto the wall in the opposite direction, the sign in expression (116) would be opposite, since the upper limit would become negative. Thus, and this clearly follows from the structure of formula (115), the sign of the slip length is determined by the choice of the reference direction in the applied coordinate system. For a gas bubble in a spherical coordinate system, the slip length is negative, as in the example discussed above, whereas for an aerosol falling in air (Fig. 5), the slip length in air in the same system is positive.

Substituting Eq. (115) in Eq. (111), one can obtain

$$-p_0'-p'+2\eta'\frac{\partial V_z'}{\partial z}=-nk_BT+\frac{2\eta'}{\lambda'}V_z'. \tag{117}$$

Derived here for the first time Eq. (117) that describes the normal component of viscous stress tensor in a gas near a liquid wall, should be considered alongside the previously known Eq. (35), describing tangential partial slippage. From the standpoint of molecular kinetic theory, Eq. (117) and Eq. (35) have the same nature.

Given the idealization of the system under consideration, it is better to use a less strict relation instead of (Eq. 116).

$$\lambda'=-kL, \tag{118}$$

where the coefficient $k$ is a positive number near 1, so that $|\lambda'|$ is of the order of the mean free path of a molecule in the gas phase $L$.

Since normal slip condition (117) assumes a non-zero gas-dynamic velocity near the interface $V_z'$, gas impermeability to a dense wall in a stationary system has to be achieved through diffusion drift. Indeed, according to Eq. (91), zero gas flow through the wall corresponds to the following condition at the interface:

$$nV_z'-D\frac{\partial n}{\partial z}=0. \tag{119}$$

Thus, due to the compressibility of the gas, its density has to be non-uniform. Taking into account Eq. (92) and Eq. (119), the expression (117) can be rewritten as

$$-p_0{}'-p'+2\eta'\frac{\partial V_z{}'}{\partial z}=-n_0k_BT-k_BT\Delta n+\frac{2\eta' D}{\lambda' n}\frac{\partial n}{\partial z}, \quad (120)$$

where the static gas pressure is determined by the formula:

$$p_0{}'=n_0k_BT. \quad (121)$$

Since the diffusion coefficient of an ideal gas is related to the dynamic viscosity by the formula

$$\eta'=mnD, \quad (122)$$

then, taking into account Eq. (105), expression (120) can be rewritten as

$$-p_0{}'-p'+2\eta'\frac{\partial V_z{}'}{\partial z}=-n_0k_BT-k_BT\Delta n+\frac{k_BT\lambda'}{\pi}\frac{\partial n}{\partial z}. \quad (123)$$

If in the reference frame under consideration the dense phase moves with velocity $V_z$, then the relative velocity at the interface has to be substituted in Eq. (116):

$$\sigma_{zz}{}'=-p_0{}'-p'+\frac{2\eta'}{\lambda'}(V_z{}'-V_z), \quad \text{where} \quad \lambda'<0. \quad (124)$$

Similarly,

$$\frac{\partial V_z{}'}{\partial z}=\frac{V_z{}'-V_z}{\lambda'}, \quad \text{where} \quad \lambda'<0. \quad (125)$$

Next, one can introduce the tangential viscous stress for the liquid at the gas interface

$$\sigma_{xz}=\frac{\eta}{\lambda}(V_x-V_x{}'), \quad (126)$$

and a similar relation for the normal component $\sigma_{zz}$.

For air under normal conditions, the mean free path $L \approx 67.3\,\mathrm{nm}$ (air humidity changes this by no more than a few percent), viscosity, and density are equal to $\eta' \approx 1.84 \cdot 10^{-5}\mathrm{Pa \cdot s}$ and $\rho' \approx 1.17\,\mathrm{kg \cdot m^{-3}}$, respectively, [21-22].

According to Eq.(116), the modulus of the slip length in the gas phase is estimated by the mean free path in air: $|\lambda'| \approx 100\,\mathrm{nm}$. Then, the estimate of the corresponding slip length in the liquid phase, assuming water, for which $\eta \sim 10^{-3}\,\mathrm{Pa \cdot s}$, is $\lambda = |\lambda'| \dfrac{\eta}{\eta'} \approx 5\,\mu\mathrm{m}$ . The slip coefficient, determined by formula (79), is estimated by $Sl \approx 0.005\,\mathrm{m^2 s \cdot kg}$ . The limitation on the Reynolds number for water (72) leads to the condition for the bubble radius $R < 50\,\mu\mathrm{m}$ .

It can be shown [Supporting material, Eq.(S.120)] that the rate of rise of a small bubble, taking into account tangential and normal slip in the gas, has the form

$$V_0 \approx \frac{(\rho' - \rho) g R^2}{3\eta} \left\{ \frac{3|\lambda'|}{R} + \frac{\eta'}{\eta} + 1 + \frac{3\lambda'^2}{\pi R^2} \right\} \left\{ \frac{3|\lambda'|}{R} + \frac{3\eta'}{2\eta} + 1 + \frac{3\lambda'^2}{\pi R^2} \right\}^{-1} . \tag{127}$$

Here, the numerator and denominator represent the transverse slip, which is linear in the slip length in the bubble, $\lambda'$, defined by Eq. (116), while the normal slip is represented by terms quadratic in this value.

Since for bubbles of the sizes under consideration $|\lambda'| R^{-1} << 1$, and given that for an air bubble under normal conditions, $\eta' \eta^{-1} \approx 1.84 \cdot 10^{-2}$, the rise velocity of a bubble is determined with high accuracy by the Hadamard-Rybczynski equation (13). The resulting contribution of tangential slip is proportional to $\lambda'^2 R^{-2}$, while that of normal slip is almost two orders of magnitude smaller, i.e., $\lambda'^2 R^{-2} \eta' \eta^{-1}$ .

In the zeroth-order expansion in powers of $\lambda' R^{-1}$ [Supporting material, Eq. (S.122)], for the gas volume concentration inside the bubble, we have

$$n = n_0 \left( 1 - \frac{\rho g r}{p_a} \cos\theta \right), \tag{128}$$

Thus, the bubble density is slightly lower at the top and slightly higher at the bottom than at the center. However, at the Reynolds numbers considered, this effect is very small.

The characteristic relative increment in air density $\frac{\rho g R}{p_a}$ for a bubble of radius $R = 10\,\mu\text{m}$ is $10^{-6}$ . This value increases with increasing bubble radius, and as the radius approaches 1 mm, it can reach a value that leads to noticeable deviations from the spherical shape of the bubble. The regime of motion of such large bubbles meets the criteria Re>1 and Pe>1, which goes beyond the limitations adopted here and requires taking into account higher-rank spherical modes and the corresponding shape deformations.

## 6. Application to the aerosol

Let us consider a small liquid droplet of density $\rho'$ that falls vertically downwards at a constant velocity $V_0$ in the still air of density $\rho$ . The droplet has a spherical shape stabilized by interfacial surface tension. The *z* axis is oriented vertically upwards (Fig. 5).

In this case, using the notation adopted here, Eq. (116) for air can be rewritten as

$$\lambda = \frac{4}{3} L , \tag{129}$$

and the slippage of the liquid inside the droplet (the hatched parameters) will be described by the negative quantity $\lambda' = -\lambda \frac{\eta'}{\eta}$. Find the velocity of the fall of a liquid droplet in the atmosphere within the same assumptions as in the previous discussion for the bubble: the small Reynolds and Peclet numbers.

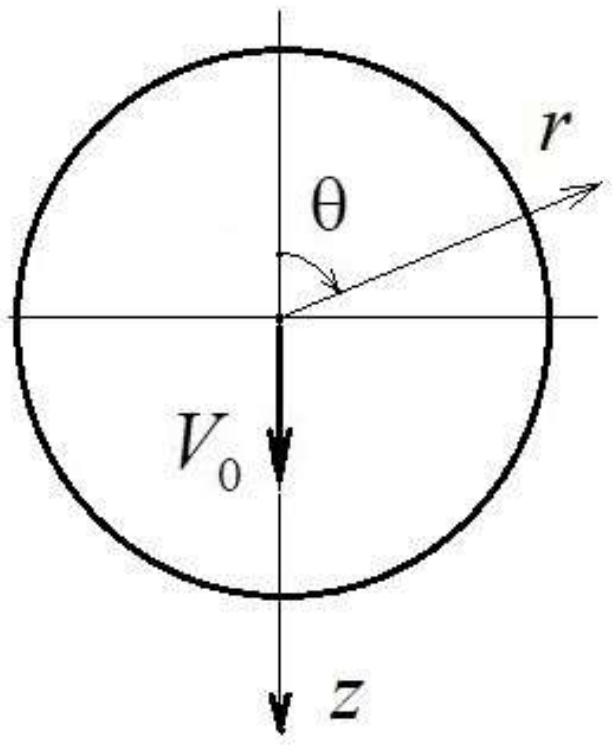

**Fig. 5.** Polar coordinates $(r, \theta)$ that include the radial coordinate and polar angle, respectively; the positive direction of the droplet's velocity $V_0$ is indicated by the arrow.

In the frame of reference associated with the droplet, the velocity of the air at $r \to \infty$ satisfies the boundary conditions

$$V_r(\infty, \theta) = V_0 \cos\theta, \tag{130}$$

$$V_\theta(\infty, \theta) = -V_0 \sin\theta. \tag{131}$$

The boundary conditions on the droplet surface are

$$V_r'(R, \theta) = 0, \tag{132}$$

where all the dashed quantities, like a radial velocity component, $V_r'$ , characterizes the liquid inside the droplet. Eq (130) means that the droplet does not change its spherical shape.

The normal slip of air on a falling liquid droplet can be described by condition (123), which in spherical coordinates takes the form:

$$\left[-p_0 - p + 2\eta \frac{\partial V_r}{\partial r}\right]_{r=R} = k_B T \left[-n_0 - \Delta n + \frac{\lambda}{\pi} \frac{\partial n}{\partial z}\right]_{r=R}, \qquad (133)$$

At small Peclet numbers, diffusion dominates over convection, so the gas concentration obeys the Laplace equation (98), the solution of which in spherical coordinates for the external problem has the form:

$$n = n_0 + \sum_{k=0}^{\infty} f_k \left(\frac{R}{r}\right)^{k+1} P_k(\cos\theta). \qquad (134)$$

where the sum on the right-hand side corresponds to $\Delta n$. Given the form of boundary conditions (130)-(131), we restrict the sum in the Eq.(134) to the term corresponding to $k$=1:

$$n = n_0 + f_1 \frac{R^2}{r^2} \cos\theta. \qquad (135)$$

according to Eq. (119), taking into account $|\Delta n| << n_0$, the boundary condition has the form

При этом, согласно (119) с учетом $|\Delta n| << n_0$ граничное условие имеет вид

$$V_r(R,\theta) = \frac{D}{n} \frac{\partial \Delta n}{\partial r}\bigg|_{r=R} \approx -\frac{2Df_1}{n_0 R} \cos\theta, \qquad (136)$$

where $f_1$ is a constant.

Let us consider the boundary conditions caused by the continuity of viscous stresses (tangential and normal components) acting at the interface of two liquids, [20,44].

The static pressure in the air on the surface of the droplet is determined by

$$p_0 = C - \rho g R \cos\theta, \qquad (137)$$

where C is a constant.

Similarly, the liquid in droplet creates the pressure:

$$p_0' = C' - \rho' g R \cos\theta, \qquad (138)$$

where $C$ is a constant.

The surface of the liquid sphere is affected by the resulting pressure equal to the difference between Eq. (138) and Eq.(137)

$$\Delta p_0(\theta) = p_0 - p_0' = (\rho' - \rho) g R \cos\theta + C - C'. \qquad (139)$$

It determines the Archimedean force acting on the droplet,

$$F_z = 2\pi R^2 \int_0^{\pi} \Delta p_0(\theta) \sin\theta \cos\theta d\theta = (\rho' - \rho) g v, \qquad (140)$$

where $v = \frac{4}{3}\pi R^3$ is the volume of the liquid droplet.

Considering that the equality of the normal components of the viscous stress tensor must be satisfied at the interface, [20,44]

$$\sigma_{rr} = \sigma_{rr}', \qquad (141)$$

i.e.

$$-(p + p_0) + 2\eta \frac{\partial V_r}{\partial r} = -(p' + p_0') + 2\eta' \frac{\partial V_r'}{\partial r} \qquad (142)$$

Eq. (142) can be conveniently rewritten taking into account Eq.(139) as

$$-p + 2\eta \frac{\partial V_r}{\partial r} + p' - 2\eta' \frac{\partial V_r'}{\partial r} = p_0 - p_0' = (\rho' - \rho) g R \cos\theta, \qquad (143)$$

where the right-hand side of the expression contains only terms proportional to $\cos\theta$.

Similarly, condition (133) takes the form

$$\left[ -p + 2\eta \frac{\partial V_r}{\partial r} \right]_{r=R} = -k_B T f_1 \left[ 1 - \frac{\lambda}{\pi R} \right]_{r=R}. \qquad (144)$$

The equality of the tangential viscous stresses of a liquid droplet and air at their interface is given by Eqs. (26)-(27). The condition of the tangential partial slip is determined by

$$\sigma_{r\theta}(\theta)=\eta\left(\frac{1}{r}\frac{\partial V_r}{\partial\theta}+\frac{\partial V_\theta}{\partial r}-\frac{V_\theta}{r}\right)_{r=R}=\frac{\eta}{\lambda}(V_\theta(R,\theta)-V_\theta'(R,\theta)) \tag{145}$$

or

$$\sigma_{r\theta}'(\theta)=\eta'\left(\frac{1}{r}\frac{\partial V_r'}{\partial\theta}+\frac{\partial V_\theta'}{\partial r}-\frac{V_\theta'}{r}\right)_{r=R}=\frac{\eta'}{\lambda'}(V_\theta'(R,\theta)-V_\theta(R,\theta)), \tag{146}$$

where λ and λ' are the slip parameters of the external and internal liquids, respectively. Eq. (145) and Eq. (146) generalize the Navier boundary condition of partial slip of a liquid over a solid. Obviously, applying this condition to the liquid-liquid interface, we must use the relative velocity of both liquids on the surface. Taking into account that there is a stitching of the tangential stresses given by Eq. (27), we have

$$\frac{\eta}{\lambda}=-\frac{\eta'}{\lambda'} \quad \text{or} \quad \frac{\lambda}{\lambda'}=-\frac{\eta}{\eta'}. \tag{147}$$

Since conditions (145) and (146) duplicate each other, we will further use only one of them, namely, the condition for the external liquid given by Eq. (145).

It can be shown [(A27), Appendix A] that, within the framework of the considered problem, the velocity of a droplet fall is described by

$$V_0=\frac{2(\rho'-\rho)gR^2}{9\eta}\left(\frac{3\lambda}{R}+1+\frac{\eta}{\eta'}+\frac{3\lambda^2}{2\pi R^2}\right)\left(\frac{2\lambda}{R}+1+\frac{2\eta}{3\eta'}+\frac{\lambda^2}{\pi R^2}\right)^{-1}, \tag{148}$$

where the terms with λ to the first power in the numerator and denominator correspond to tangential slip at the interface, and the terms containing $\lambda^2$ correspond to normal slip. In Eq.(148), the fraction before the parentheses corresponds to the Stokes equation (11), and the remainder is a factor that differs little from 1:

$$\left(\frac{3\lambda}{R}+1+\frac{\eta}{\eta'}+\frac{3\lambda^2}{2\pi R^2}\right)\left(\frac{2\lambda}{R}+1+\frac{2\eta}{3\eta'}+\frac{\lambda^2}{\pi R^2}\right)^{-1}\approx 1+\frac{\eta}{3\eta'}+\frac{\lambda}{R}-\frac{(4\pi-1)\lambda^2}{2\pi R^2}+\ldots \tag{149}$$

If the medium surrounding the drop is also a liquid, then there is no normal slip, so the terms with $\lambda^2$ must be excluded in Eq. (148), and, taking into account Eqs.(30), we obtain the well-known formula (7) with $\lambda = \beta R$ [or its variants given by Eqs. (60), (70), (S.78)], describing the tangential slip of a liquid drop in an external liquid.

For a water droplet falling in air under normal conditions, Eq. (72) is satisfied if $R << 35\ \mu\text{m}$. Experimental data corresponding to the terminal velocity of a microdroplet falling in air, having a radius in the range $0.25 \le R \le 9.5\,\mu\text{m}$ (or $0.07 < \text{Kn} < 0.3$ ), satisfying the conditions of the problem under consideration, are generalized in the Ref. [31] by following semi-empirical relation:

$$V_{\exp}(x) = \frac{2(\rho' - \rho)gR^2}{9\eta}\left\{1 + x^{-1}\left[1.257 + 0.400\exp\left(-1.10x\right)\right]\right\}, \tag{150}$$

where $x = RL^{-1} = \text{Kn}^{-1}$ is the ratio of the droplet radius $R$ to the mean free path in air $L \approx 70\,\text{nm}$.

The Cunningham–Millican correction factor included in Eq. (150) when substituting the Knudsen number takes the traditional form (see, for example, Ref. [5], p. 258):

$$C_c = 1 + \text{Kn}\left[1.257 + 0.400\exp\left(-1.10\ \text{Kn}^{-1}\right)\right]. \tag{151}$$

It is noted that the error in this range can reach 7.4%., Ref. [31].

Recalculating the slip length $\lambda$ through the mean free path in air using Eq. (129), we rewrite Eq. (148) as a function of $x$.

$$V_0(x) = \frac{2(\rho' - \rho)gR^2}{9\eta}\left(1 + \frac{\eta}{\eta'} + \frac{4}{x} + \frac{8}{3\pi x^2}\right)\left(1 + \frac{2\eta}{3\eta'} + \frac{8}{3x} + \frac{16}{9\pi x^2}\right)^{-1}. \tag{152}$$

It is obvious that the correction factor included in formula (151), expressed through the Knudsen number, has the form

$$C' = \left(1+\frac{\eta}{\eta'}+4\mathrm{Kn}+\frac{8}{3\pi}\mathrm{Kn}^2\right)\left(1+\frac{2\eta}{3\eta'}+\frac{8}{3}\mathrm{Kn}+\frac{16}{9\pi}\mathrm{Kn}^2\right)^{-1}. \qquad (153)$$

Fig. 6 shows a comparison of the correction functions corresponding to Eqs. (151) and (153) in the range of the dimensionless parameter $3.7 \le x \le 136$ (or $0.07 < \mathrm{Kn} < 0.27$)

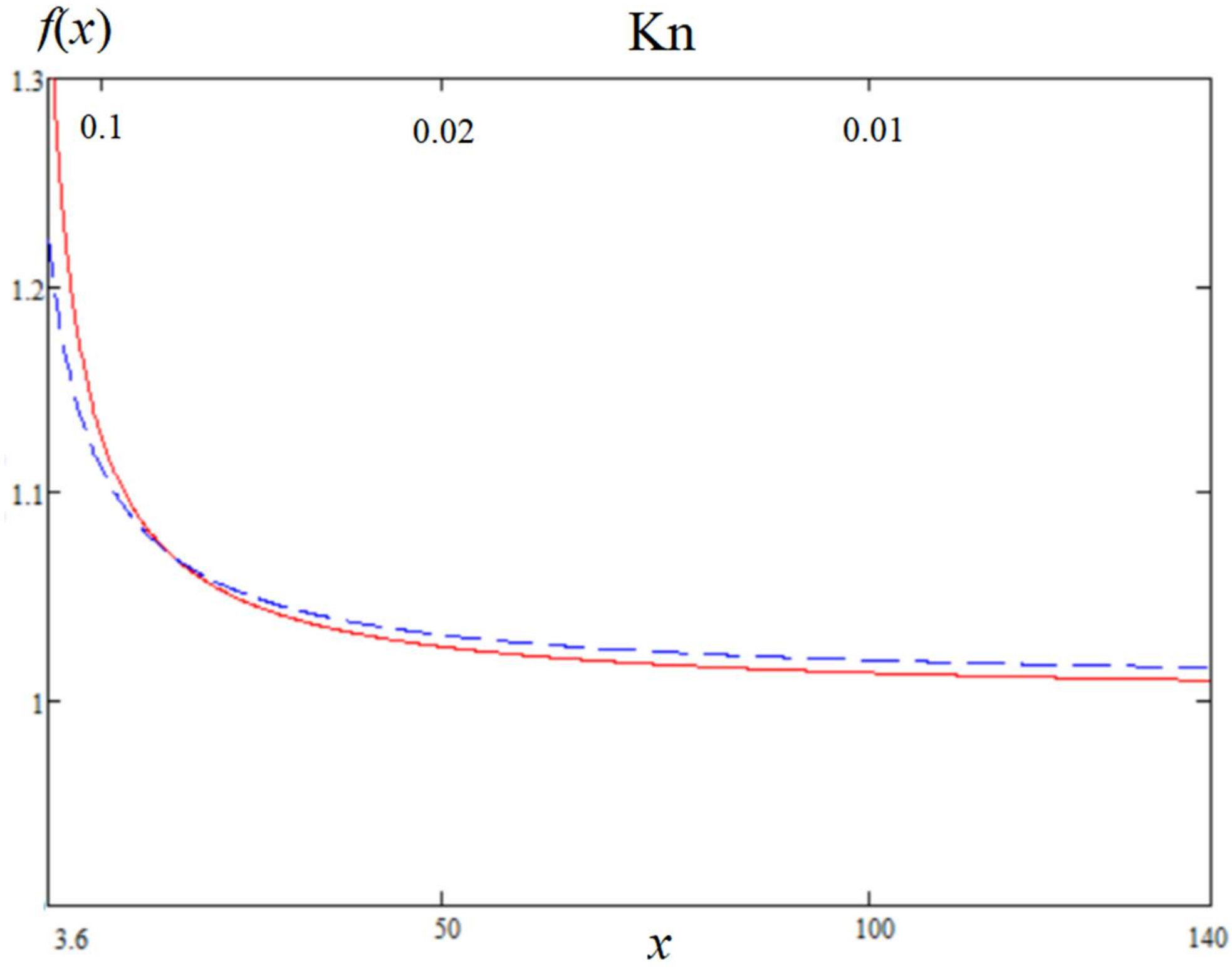

**Fig. 6.** Graphs of the dimensionless functions $f(x) = 9\eta V_0(x)[2(\rho'-\rho)gR^2]^{-1}$ corresponding to: red solid line is the Conningham–Millican factor $C_c$, Eq. (151), blue dotted line is the theoretical approach found according to Eq. (153); $x = RL^{-1} = \mathrm{Kn}^{-1}$ is dimensionless radius of the droplet.

The relative error of Eq.(153) in the region of its greatest deviation from Eq. (151) at $x = 3.7$ is approximately 10%, i.e., is approximately at the level of experimental error. Note that accounting for normal slip improves the convergence of theory with experiment at the smallest droplet radii, when this contribution is greatest. However, the relative contribution of normal slip is estimated by $\pi^{-1}x^{-2}$ in the considered range of droplet sizes and the flow regime (Re<<1, Pe<<1), and does not exceed 1%.

We see that factors (151) and (153) are nearly identical for $x$>10 (Kn<0.1). Taking into account the different methods of obtaining them, that correspondence provides a validation for the approach proposed here.

Let us determine the deviation of the air concentration near the droplet corresponding to the presence of normal slip. Expressing $f_1$ from Eq.(A22), using Eqs. (A26)-(A27), in the zeroth-order expansion in $\lambda R^{-1}$, one can obtain

$$f_1 \approx -\frac{\rho' g R}{3k_B T}, \tag{154}$$

where $\rho'$ is the density of the liquid in the falling droplet. Substituting Eq. (154) into Eq. (135), we find a solution for the concentration of air molecules near the droplet

$$n = n_0\left(1 - \frac{\rho' g R^3}{3 p_a r^2}\cos\theta\right), \tag{155}$$

where $p_a = n_0 k_B T$ is the atmospheric pressure.

Considering Fig. 5, equation (155) suggests that in the air in contact with the falling droplet along its forward motion, the concentration of molecules is increased, and the air is denser, while in the air opposite the droplet motion, the density of the air is reduced. This corresponds to basic ideas about air flow around a falling spherical body.

The characteristic relative air compaction $\frac{\rho' g R}{3 p_a}$ for a droplet with radius $R = 3\,\mu\text{m}$ is $10^{-7}$. This value increases with the droplet radius, and as it approaches 1 mm, it can reach a value that leads to noticeable deviation from the spherical shape of the droplet. However, the regime of such large drop falling meets the criteria Re>1 and Pe>1, which exceeds the limitations adopted here and requires consideration of higher-rank spherical modes and their corresponding shape deformations.

## 7. Outlook

A solution to the problem of low-Reynolds-number flow around a spherical fluid droplet moving in another fluid is proposed. A generalization of the HRE is presented that takes into account partial tangential and normal slip at the interface at moderate Knudsen numbers (Kn<0.3). Normal slip is accompanied by the density gradient in the fluid and is applicable only if one of the phases in contact at the interface is a gas.

Although tangential partial slip and the associated generalization of the HRE have been considered previously [34-36], they were done using the friction coefficient formalism. Here, this issue is discussed within the more general formalism of two slip lengths. It is proven that each of the two fluids separated by an interface has its own slip length. There is a relationship between the pair of slip lengths of the contacting fluids.

The existence of two slip lengths follows from the equality of the fluids that form the interface. If the concept of slip length is introduced, a pair of slip lengths automatically arises. In the case of steady flow, they are related by the value of the viscosity ratio taken with a minus sign., i.e., they have opposite signs.

The choice of sign, generally speaking, is not absolute but is related to the choice of coordinate axes, i.e., it depends on the coordinate system. However, if the choice of a specific coordinate system has already been made, the signs of the two slip lengths are determined unambiguously, namely, in such a way that slip at the interface occurs in the direction of the applied external force. In this case, the power released during friction is positive, which corresponds to the heat release at the interface.

In this case, the leading fluid at the interface moves faster than the trailing fluid and drags it along in the direction of the external force. By leading fluid, we mean the fluid that is directly driven by the external force. For example, if we consider the motion of a liquid droplet in an external fluid, the leading fluid is the external fluid, as it directly interacts with the vessel walls or the boundary conditions at infinity, which are the hydrodynamic analog of a thermostat.

The fluid inside the droplet interacts with the boundary conditions at infinity through the external fluid and is therefore the trailing fluid. Considering the velocities of both fluids at the interface from this point of view allows to unambiguously determine, within the adopted reference frame, for which fluid the slip length is positive and for which it is negative. The choice of signs for the slip lengths is demonstrated using the example of a plane flow, as well as the example of a flow corresponding to the motion of a spherical liquid droplet in an external liquid.

Such a deep mutual dependence of the slip lengths of two contacting liquids allows us to speak not simply of the twoness of these lengths, which is trivial, but of the deep duality of their physical nature.

Developing a scientific foundation for methods for calculating the friction coefficient or the corresponding slip length at the interface between two fluids is an actual problem in fluid dynamics, generally unsolved. However, if a dense phase and a gas are in contact at the interface, methods of the molecular kinetic theory of gases allow for well-founded estimates of the slip length.

The so-called Maxwellian boundary condition, defined on the surface y=0, in the approximation of complete, or diffuse accommodation has the form [47]:

$$\left|V_x'(y=0)-V_x(y=0)\right| = L\left|\frac{\partial V_x'}{\partial y}\right|_{y=0}, \tag{156}$$

where $V_x'$ is the gas velocity; $V_x$ is the velocity of the dense phase (liquid or solid), $L$ is the mean free path of the gas. From relation (156), it follows that the slip length is related to the mean free path by the relation:

$$\lambda = L. \tag{157}$$

In Section 5 of this paper, a slightly different relation, defined by Eq. (116) or Eq. (129), is derived using the molecular kinetic approach in the diffusion limit for an ideal gas, namely $\lambda = \frac{4}{3}L$. This result was obtained here by investigation the transfer of momentum at the gas-liquid interface through the tangential and normal components of the viscous stress tensor.

When examining the interaction of a gas molecule with a liquid surface, it was assumed that after colliding with the liquid surface, the gas molecules transfer all their momentum to the liquid and then return to the gas phase with a momentum whose value is statistically distributed corresponding to the Maxwell distribution at the interface temperature. This allows to determine the resulting momentum transfer. The effective slip length is determined from the continuity of the viscous stress tensor components at the interface.

We emphasize that the approach to momentum transfer at the gas-liquid interface used here differs significantly from both the approach originally introduced by Cunningham [25] and methods based on kinetic equations [10,50]. Our approach is based on the requirement of continuity of the normal and tangential components of the viscous stress tensor at the gas-liquid interface that allows for a natural description of slip in such a system using the Stokes hydrodynamic equations. Within this approach, the HRE can be generalized to adequately describe systems with a gas-liquid interface (bubbles and aerosols).

It is common to introduce momentum accommodation coefficients—tangential (TMAC) and normal (NMAC)—to fit the slip length. The proposed approach eliminates the need to introduce these coefficients when considering the momentum transfer at a gas-liquid interface, since the slip length is calculated directly from theory. Introducing these accommodation coefficients is indeed necessary when considering a gas-solid interface. In this case, the viscous stress tensor terminates at the interface, so it is necessary to carefully consider the interaction of the gas molecule with the surface, taking into account many dynamic and kinetic factors.

In Section 6, in the limit of small Reynolds numbers and moderate Knudsen numbers, an expression is derived for the small liquid droplet terminal velocity in still air. This expression fits the experimental data well when using the slip length defined by formula (129). Expression (157) yields significantly poorer agreement with experiment. Taking normal slip into account significantly improves agreement with experiment at small droplet radii, when the contribution of normal slip is more noticeable. Figure 6 shows a good agreement between the theoretical result and the semi-empirical Cunningham–Millikan expression that fits the experimental data at the moderate Knudsen numbers, $Kn<0.1$.

It is currently accepted that the HRE should describe the system under consideration well. The existing deviations, according to Levich [20] and many other researchers [23,48], should be associated with the difficulty of ensuring sufficient purity of liquids, and unaccounted surfactants distort the interpretation of experimental result. There is another previously unaccounted possibility for interpreting such deviations, related to the mechanism of partial slip at the liquid-liquid interface based on the Navier boundary condition. The question of how much the partial slip mechanism manifests itself in every system with a liquid-liquid interface can be also resolved experimentally.

First experiment that can be proposed in this connection is the following. Consider a system of two immiscible liquids A and B. The first experiment consists of measuring the velocity $V_A$ of a small spherical droplet of liquid A with radius *R* in a large reservoir filled with liquid B. Then, if the motion of the droplet is determined by the condition of partial slip, the slip length, according to (60), is determined by:

$$\lambda_B = \frac{(\rho_B - \rho_A) g R^2 - 3\eta_B V_A}{6(\rho_B - \rho_A) g R} - R\left(\frac{\eta_B}{3\eta_A} + \frac{1}{2}\right). \tag{158}$$

In the second experiment, the liquids A and B are swapped, so that the velocity $V_B$ of a droplet of liquid B of the same radius *R* is measured in a large reservoir filled with liquid A. Then, instead of (158), we can write

$$\lambda_A = \frac{(\rho_A - \rho_B) g R^2 - 3\eta_A V_B}{6(\rho_A - \rho_B) g R} - R\left(\frac{\eta_A}{3\eta_B} + \frac{1}{2}\right). \tag{159}$$

If the partial slip mechanism really operates in the liquid-liquid system under consideration, relation (30) must be satisfied, namely:

$$\frac{\lambda_B}{\lambda_A} = -\frac{\eta_B}{\eta_A}. \tag{160}$$

If there is a deviation from Eq. (160), that indicates a noticeable influence of other mechanisms of interaction between liquids on the interface.

The field of application of the generalized HRE is the motion of hydrophobic (lipophilic) liquids in water and vice versa, i.e. the description of stratification and sedimentation of aqueous emulsions or vice versa – water droplets in oils, etc. Presumably, the best applicability of this equation should be expected for the interface of hydrophobic liquid and hydrophilic one (water – hydrocarbons, water – higher alcohols, in general: aqueous emulsions, water – lipophilic organic liquids and oils, etc.). These are quite important emulsions in practical terms, for example, for the oil industry and medicine. The use of such a parameter as slip length allows the experiment and theory to be reconciled.

**Data availability statement**. The data that support the findings of this study are available from the author upon reasonable request.

**Appendix A**

The general solution of the Stokes equations given by Eqs. (8)-(9) for an axisymmetric problem in spherical coordinates given in the Table 1 was presented in [49]. The radial and polar components of the incompressible fluid velocity, pressure and stream function are shown in the rows from top to bottom of the Table; $P_l(\cos\theta)$ is the Legendre polynomial, $P_l^1(\cos\theta)$ is the associated Legendre function of the first order.

**Table 1. General solution of internal and external axisymmetric problems of the Stokes equations in a spherical coordinate system**

| Internal problem ( $r \le R$ ) | External problem ( $r \ge R$ ) |
|---|---|
| $V_r(r,\theta)=\sum_{l=1}^{\infty} l(l+1)\left\{\frac{a_l}{4l+6}\left(\frac{r}{R}\right)^{l+1}+c_l\left(\frac{r}{R}\right)^{l-1}\right\}P_l(\cos\theta)$ | $V_r(r,\theta)=-\sum_{l=1}^{\infty} l(l+1)\left\{\frac{b_l}{4l-2}\left(\frac{R}{r}\right)^{l}+d_l\left(\frac{R}{r}\right)^{l+2}\right\}P_l(\cos\theta)+d_0\left(\frac{R}{r}\right)^2$ |
| $V_\theta(r,\theta)=\sum_{l=1}^{\infty}\left\{a_l\frac{l+3}{4l+6}\left(\frac{r}{R}\right)^{l+1}+c_l(l+1)\left(\frac{r}{R}\right)^{l-1}\right\}P_l^1(\cos\theta)$ | $V_\theta(r,\theta)=\sum_{l=1}^{\infty}\left\{b_l\frac{l-2}{4l-2}\left(\frac{R}{r}\right)^{l}+d_l l\left(\frac{R}{r}\right)^{l+2}\right\}P_l^1(\cos\theta)$ |
| $p(r,\theta)=\frac{\eta}{R}\sum_{l=0}^{\infty}(l+1)a_l\left(\frac{r}{R}\right)^{l}P_l(\cos\theta)$ | $p(r,\theta)=-\frac{\eta}{R}\sum_{l=1}^{\infty} l b_l\left(\frac{R}{r}\right)^{l+1}P_l(\cos\theta)$ |
| $\Psi(r,\theta)=-R^2\sin\theta\sum_{l=1}^{\infty}\left\{\frac{a_l}{4l+6}\left(\frac{r}{R}\right)^{l+3}+c_l\left(\frac{r}{R}\right)^{l+1}\right\}P_l^1(\cos\theta)$ | $\Psi(r,\theta)=R^2\sin\theta\sum_{l=1}^{\infty}\left\{\frac{b_l}{4l-2}\left(\frac{R}{r}\right)^{l-2}+d_l\left(\frac{R}{r}\right)^{l}\right\}P_l^1(\cos\theta)-d_0R^2\cos\theta$ |

Taking this into account, based on the form of boundary conditions (130)-(131), the desired solution for the velocity of the fluid outside the droplet can be represented in the form

$$V_r=\left\{V_0-b\frac{R}{r}-2d\left(\frac{R}{r}\right)^3\right\}\cos\theta, \tag{A1}$$

$$V_\theta=\left\{\frac{bR}{2r}-d\left(\frac{R}{r}\right)^3-V_0\right\}\sin\theta, \tag{A2}$$

$$p=-\frac{\eta}{R}b\left(\frac{R}{r}\right)^2\cos\theta. \tag{A3}$$

Substituting Eq. (136) in Eq. (A1), we have

$$b=V_0-2d+\frac{2Df_1}{n_0R}. \tag{A4}$$

Substituting Eq. (A4) into Eqs. (A1)-(A3), we obtain

$$V_r=\left\{V_0-\left(V_0-2d+\frac{2Df_1}{n_0R}\right)\frac{R}{r}-2d\left(\frac{R}{r}\right)^3\right\}\cos\theta, \tag{A5}$$

$$V_\theta=\left\{\left(V_0-2d+\frac{2Df_1}{n_0R}\right)\frac{R}{2r}-d\left(\frac{R}{r}\right)^3-V_0\right\}\sin\theta, \tag{A6}$$

$$p=-\frac{\eta}{R}\left(V_0-2d+\frac{2Df_1}{n_0R}\right)\left(\frac{R}{r}\right)^2\cos\theta. \tag{A7}$$

Obviously, solution for flow inside a droplet have to be described only by the solution of the internal problem

$$V_r'=2\left\{\frac{a'}{10}\left(\frac{r}{R}\right)^2+c'\right\}\cos\theta, \tag{A8}$$

$$V_\theta'=-\left\{a'\frac{4}{10}\left(\frac{r}{R}\right)^2+2c'\right\}\sin\theta. \tag{A9}$$

$$p'=p_0'+\frac{2a'\eta' r}{R^2}\cos\theta. \tag{A10}$$

Taking into account the condition (132), for the radial component of the velocity of the internal fluid (A8) we have

$$a'=-10c'. \tag{A11}$$

Then, the system of equations for the internal fluid (A8)-(A10) can be rewritten as

$$V_r' = 2c'\left\{1-\left(\frac{r}{R}\right)^2\right\}\cos\theta, \tag{A12}$$

$$V_\theta' = 2c'\left\{2\left(\frac{r}{R}\right)^2 - 1\right\}\sin\theta, \tag{A13}$$

$$p' = -\frac{20c'\eta' r}{R^2}\cos\theta. \tag{A14}$$

Substituting (A6)-(A7) and (A12)-(A13) in condition (27), we obtain

$$c' = \frac{\eta}{\eta'}d\,. \tag{A15}$$

Then, the Eqs. (A12)-(A13) can be rewritten as

$$V_r' = 2\frac{\eta}{\eta'}d\left\{1-\left(\frac{r}{R}\right)^2\right\}\cos\theta, \tag{A16}$$

$$V_\theta' = 2\frac{\eta}{\eta'}d\left\{2\left(\frac{r}{R}\right)^2 - 1\right\}\sin\theta. \tag{A17}$$

$$p' = -\frac{\eta}{\eta'}d\,\frac{20\eta' r}{R^2}\cos\theta = -d\,\frac{20\eta r}{R^2}\cos\theta. \tag{A18}$$

To use the boundary condition (143), evaluate the expressions

$$-p(R,\theta) + 2\eta\left.\frac{\partial V_r(r,\theta)}{\partial r}\right|_R = \frac{3\eta}{R}(V_0 + 2d)\cos\theta + \frac{6\eta D f_1}{n_0 R^2}\cos\theta, \tag{A19}$$

$$-p'(R,\theta) + 2\eta'\left.\frac{\partial V_r'(r,\theta)}{\partial r}\right|_R = \frac{12 d\eta}{R}\cos\theta. \tag{A20}$$

Substituting Eqs. (A19) and (A20) into Eq. (143), one can obtain

$$\frac{3\eta}{R}(V_0 + 2d)\cos\theta + \frac{6\eta D f_1}{n_0 R^2} - \frac{12 d\eta}{R}\cos\theta = (\rho' - \rho) g R\cos\theta. \tag{A21}$$

On the other hand, using the normal slip condition (133), we obtain

$$\frac{3\eta}{R}(V_0+2d)=-k_B T f_1\left[1-\frac{\lambda}{\pi R}+\frac{3\lambda^2}{\pi R^2}\right]. \qquad \text{(A22)}$$

In deriving Eq. (A22), it was taken into account that, according to Eqs. (10), (116), and (122), the following relation holds for the air outside the falling droplet

$$\eta D=\frac{\lambda^2}{2\pi}nk_B T\,. \qquad \text{(A23)}$$

Expressing $f_1$ from Eq. (A22) and substituting it into Eq. (A21), we obtain, writing out terms up to the order $\lambda^2 R^{-2}$

$$V_0-2d-\frac{3\lambda^2}{\pi R^2}(V_0+2d)=\frac{(\rho'-\rho)gR^2}{3\eta}. \qquad \text{(A24)}$$

Finally, using the normal slip condition (145), we have

$$\lambda\frac{6d}{R}\sin\theta=\left(-\frac{V_0}{2}-2d+\frac{Df_1}{n_0 R}-2\frac{\eta}{\eta'}d\right)\sin\theta\,, \qquad \text{(A25)}$$

where, taking into account (A22), we find

$$d=-\frac{V_0}{4}\left(1+\frac{3\lambda^2}{\pi R^2}\right)\left(\frac{3\lambda}{R}+\frac{\eta}{\eta'}+1+\frac{3\lambda^2}{2\pi R^2}\right)^{-1}. \qquad \text{(A26)}$$

Substituting Eq.(A26) into Eq. (A24), one can obtain

$$V_0=\frac{2(\rho'-\rho)gR^2}{9\eta}\left(\frac{3\lambda}{R}+1+\frac{\eta}{\eta'}+\frac{3\lambda^2}{2\pi R^2}\right)\left(\frac{2\lambda}{R}+1+\frac{2\eta}{3\eta'}+\frac{\lambda^2}{\pi R^2}\right)^{-1}. \qquad \text{(A27)}$$

## Supporting material

# Tangential and normal partial slip at the liquid-fluid interfaces: application to a small liquid droplet, gas bubble, and aerosol

Peter Lebedev-Stepanov
Shubnikov Institute of Crystallography, Kurchatov Complex of Crystallography and Photonics, Leninskiy Prospekt 59, Moscow 119333, Russia
E-mail: lebstep.p@crys.ras.ru
Author ORCID https://orcid.org/0000-0002-7009-4319

### 1. Boundary conditions

Let us consider a liquid droplet placed inside another liquid. Both liquids are insoluble in each other, do not mix with each other, and have a clear interface. The drop has a spherical shape stabilized by interfacial surface tension. The *z* axis is oriented vertically upwards (Fig. S1). If the drop density $\rho'$ is less than the liquid density $\rho$, it floats vertically upwards with a steady-state velocity $V_0$.

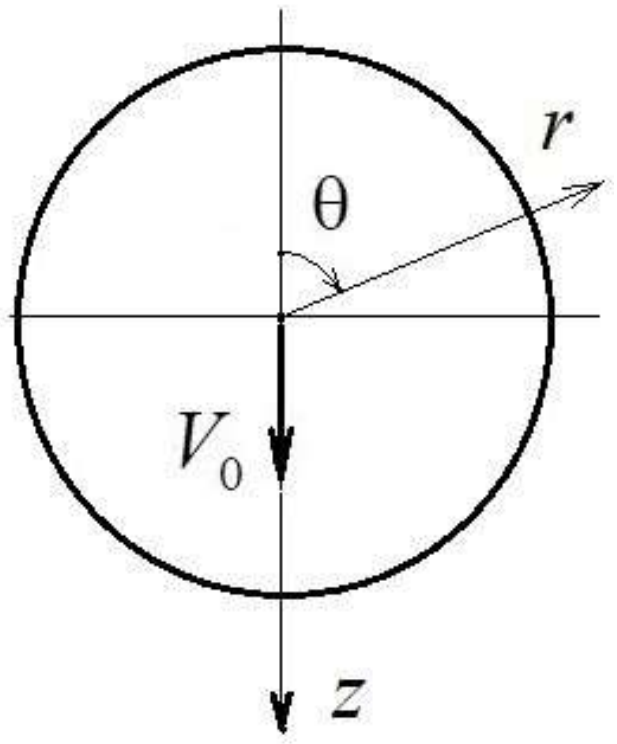

FIG S1. Polar coordinates $(r,\theta)$ that include the radial coordinate and polar angle, respectively; the positive direction of the droplet's velocity $V_0$ is indicated by the arrow.

In the frame of reference associated with the droplet, the velocity of the external liquid at $r \to \infty$ satisfies the boundary conditions

$$V_r(\infty,\theta) = V_0 \cos\theta, \quad \text{(S.1)}$$

$$V_\theta(\infty,\theta) = -V_0 \sin\theta. \quad \text{(S.2)}$$

The boundary conditions on the droplet surface are:

$$V_r(R,\theta) = 0, \quad \text{(S.3)}$$

which means that the droplet does not change its spherical shape.
The continuity of viscous stresses (tangential and normal components) acting on the interface between two fluids is used as boundary conditions, Ref. [S.1-S2]:

$$\sigma_{r\theta} = \sigma_{r\theta}', \quad \text{(S.4)}$$

i.e.

$$\eta\left(\frac{1}{r}\frac{\partial V_r}{\partial\theta}+\frac{\partial V_\theta}{\partial r}-\frac{V_\theta}{r}\right)=\eta'\left(\frac{1}{r}\frac{\partial V_r'}{\partial\theta}+\frac{\partial V_\theta'}{\partial r}-\frac{V_\theta'}{r}\right), \tag{S.5}$$

where all the dashed quantities, like a radial velocity component, $V_r'$, characterizes the liquid inside the droplet;

$$\sigma_{rr}=\sigma_{rr}', \tag{S.6}$$

which in expanded form means

$$-(p+p_0)+2\eta\frac{\partial V_r}{\partial r}=-(p'+p_0')+2\eta'\frac{\partial V_r'}{\partial r} \tag{S.7}$$

where *p* and *p'* denote the hydrodynamic pressures in the external and internal liquids, respectively.

The hydrostatic pressure in the external liquid on the surface of the drop is determined by

$$p_0=\rho g(h-R\cos\theta), \tag{S.8}$$

where *h* is the depth of immersion of the center of the drop. The minus sign before the expression on the right takes into account that the hydrostatic pressure is compressive.

Similarly, the internal liquid creates pressure:

$$p_0'=C-\rho' gR\cos\theta, \tag{S.9}$$

where *C* is a constant.

The surface of the liquid sphere is affected by the resulting pressure equal to the difference between Eq. (S.9) and Eq.(S.8)

$$\Delta p_0(\theta)=p_0-p_0'=(\rho'-\rho)gR\cos\theta, \tag{S.10}$$

where we have neglected the terms independent of the polar angle. It determines the Archimedean force acting on the drop,

$$F_z=2\pi R^2\int_0^\pi \Delta p_0(\theta)\sin\theta\cos\theta d\theta=(\rho'-\rho)gv, \tag{S.11}$$

where $v=\frac{4}{3}\pi R^3$ is the volume of the liquid sphere.

Eq. (S.7) can be conveniently rewritten taking into account Eq. (S.10) as

$$-p+2\eta\frac{\partial V_r}{\partial r}+p'-2\eta'\frac{\partial V_r'}{\partial r}=p_0-p_0'=(\rho'-\rho)gR\cos\theta. \tag{S.12}$$

Conditions (S.1)-(S.3), (S.5), and (S.12) are common to all problems of motion in an external fluid, including the derivation of the Hadamard–Rybczynski equation (HRE). If we consider the sliding of a liquid droplet in an external fluid, it is necessary to add the condition of immobility of the internal fluid at the interface in the droplet's rest frame:

$$V_r'(R,\theta)=0, \tag{S.13}$$

as well as one of the sliding conditions. To describe the motion of liquids at the interface, we introduce the condition of the transverse (tangential) partial slip of the surface of one liquid over the surface of another, which has the form

$$\sigma_{r\theta}(\theta)=\eta\left(\frac{1}{r}\frac{\partial V_r}{\partial\theta}+\frac{\partial V_\theta}{\partial r}-\frac{V_\theta}{r}\right)_{r=R}=\frac{\eta}{\lambda}(V_\theta(R,\theta)-V_\theta'(R,\theta)) \tag{S.14}$$

or

$$\sigma_{r\theta}'(\theta) = \eta'\left(\frac{1}{r}\frac{\partial V_r'}{\partial\theta} + \frac{\partial V_\theta'}{\partial r} - \frac{V_\theta'}{r}\right)_{r=R} = \frac{\eta'}{\lambda'}(V_\theta'(R,\theta) - V_\theta(R,\theta)), \tag{S.15}$$

where λ and λ' are the slip parameters of the external and internal liquids, respectively. Eqs. (S.14) and (S.15) generalize the Navier boundary condition of partial slip of a liquid over a solid. Obviously, applying this condition to the liquid-liquid interface, we must use the relative velocity of both liquids on the surface. Taking into account that there is a stitching of the tangential stresses given by Eq. (S.4), we have

$$\frac{\eta}{\lambda} = -\frac{\eta'}{\lambda'} \quad \text{or} \quad \frac{\lambda}{\lambda'} = -\frac{\eta}{\eta'}. \tag{S.16}$$

Since conditions (S.14) and (S.15) duplicate each other, we will further use only one of them, namely, the condition for the external liquid given by Eq. (S.14).

If we consider the motion of a bubble in an external fluid, conditions (S.14) and (S.15) remain valid, but condition (S.13) is inapplicable due to the presence of longitudinal (normal) slip. The longitudinal slip can be taken into account using condition (123)[1]:

$$\left[-p_0' - p' + 2\eta'\frac{\partial V_z'}{\partial r}\right]_{r=R} = \left[-n_0 k_B T - k_B T\Delta n + \frac{k_B T\lambda'}{\pi}\frac{\partial n}{\partial r}\right]_{r=R}. \tag{S.17}$$

At small Peclet numbers, diffusion dominates over convection, so the concentration obeys the Laplace equation

$$\nabla^2 n = 0, \tag{S.18}$$

the solution of which in spherical coordinates is

$$n = n_0 + \sum_{k=1}^{\infty} f_k\left(\frac{r}{R}\right)^k P_k(\cos\theta). \tag{S.19}$$

where the sum on the right-hand side corresponds to $\Delta n$. Given the form of boundary conditions (S.1)-(S.2), we restrict ourselves in (S.19) to the term in the sum corresponding to $k$=1:

$$n = n_0 + f_1\frac{r}{R}\cos\theta. \tag{S.20}$$

Moreover, according to (119)[2] taking into account $|\Delta n| << n_0$ the boundary condition at the interface has the form

$$V_r'(R,\theta) = \frac{D}{n}\frac{\partial\Delta n}{\partial r}\bigg|_{r=R} \approx \frac{D}{n_0}\frac{\partial\Delta n}{\partial r}\bigg|_{r=R}. \tag{S.21}$$

Substituting (S.20) into (S.21), we obtain

---

[1] Reference to equation (123) in the main part of the article.

[2] Reference to equation (119) in the main part of the article.

$$V_r'(R,\theta) = \frac{Df_1}{n_0 R}\cos\theta . \qquad \text{(S.22)}$$

where $f_1$ is a constant.

## 2. Liquid-liquid interface

Here we are dealing with axisymmetric boundary conditions. Thus, we have to solve an axisymmetric problem. The general solution of the axisymmetric problem for the Stokes equations in a spherical coordinate system is presented in Table S1. The derivation of the formulae is given in Ref. [S3].

We see that the boundary conditions (S.1)-(S.2) and (S.12) require solutions containing $\cos\theta$ and $\sin\theta$. The solutions for $V_r$ and $p$ have the form of a series expansion in Legendre polynomials, and the solution for $V_\theta$ is represented by a series in the associated Legendre polynomial $P_l^1$. The orthogonality of the polynomials suggests that in order to satisfy the boundary conditions, it is necessary to limit ourselves to the terms of the series that contain $\cos\theta$ or $\sin\theta$. The remaining terms must be discarded, since the boundary conditions do not allow nonzero solutions corresponding to these excess terms to exist. Thus, in the solutions under consideration, we leave only the terms with $l$=1.

Table S1. Solutions of internal and external axisymmetric problems in a spherical coordinate system obtained in the representation of a vector potential. The radial and polar components of the incompressible fluid velocity, pressure and stream function are shown in the rows from top to bottom; $P_l(\cos\theta)$ is the Legendre polynomial, $P_l^1(\cos\theta)$ is the associated Legendre function of the first order.

| Internal problem | External problem |
|---|---|
| $V_r(r,\theta) = \sum_{l=1}^{\infty} l(l+1)\left\{\frac{a_l}{4l+6}\left(\frac{r}{R}\right)^{l+1} + c_l\left(\frac{r}{R}\right)^{l-1}\right\}P_l(\cos\theta)$ | $V_r(r,\theta) = -\sum_{l=1}^{\infty} l(l+1)\left\{\frac{b_l}{4l-2}\left(\frac{R}{r}\right)^{l} + d_l\left(\frac{R}{r}\right)^{l+2}\right\}P_l(\cos\theta) + d_0\left(\frac{R}{r}\right)^2$ |
| $V_\theta(r,\theta) = \sum_{l=1}^{\infty}\left\{a_l\frac{l+3}{4l+6}\left(\frac{r}{R}\right)^{l+1} + c_l(l+1)\left(\frac{r}{R}\right)^{l-1}\right\}P_l^1(\cos\theta)$ | $V_\theta(r,\theta) = \sum_{l=1}^{\infty}\left\{b_l\frac{l-2}{4l-2}\left(\frac{R}{r}\right)^{l} + d_l l\left(\frac{R}{r}\right)^{l+2}\right\}P_l^1(\cos\theta)$ |
| $p(r,\theta) = \frac{\eta}{R}\sum_{l=0}^{\infty}(l+1)a_l\left(\frac{r}{R}\right)^{l} P_l(\cos\theta)$ | $p(r,\theta) = -\frac{\eta}{R}\sum_{l=1}^{\infty} l b_l\left(\frac{R}{r}\right)^{l+1} P_l(\cos\theta)$ |
| $\Psi(r,\theta) = -R^2\sin\theta\sum_{l=1}^{\infty}\left\{\frac{a_l}{4l+6}\left(\frac{r}{R}\right)^{l+3} + c_l\left(\frac{r}{R}\right)^{l+1}\right\}P_l^1(\cos\theta)$ | $\Psi(r,\theta) = R^2\sin\theta\sum_{l=1}^{\infty}\left\{\frac{b_l}{4l-2}\left(\frac{R}{r}\right)^{l-2} + d_l\left(\frac{R}{r}\right)^{l}\right\}P_l^1(\cos\theta) - d_0 R^2\cos\theta$ |

For the internal problem, we have

$$V_r = 2\left\{\frac{a}{10}\left(\frac{r}{R}\right)^2 + c\right\}P_1, \qquad \text{(S.23)}$$

$$V_\theta = \left\{a\frac{4}{10}\left(\frac{r}{R}\right)^2 + 2c\right\}P_1^1 . \qquad \text{(S.24)}$$

$$p = p_0 + \frac{2a\eta r}{R^2} P_1 . \qquad \text{(S.25)}$$

For the external problem, we obtain

$$V_r = -2\left\{\frac{b}{2}\frac{R}{r} + d\left(\frac{R}{r}\right)^3\right\}\cos\theta , \qquad \text{(S.26)}$$

$$V_\theta = \left\{\frac{bR}{2r} - d\left(\frac{R}{r}\right)^3\right\}\sin\theta , \qquad \text{(S.27)}$$

$$p = -\frac{\eta}{R} b\left(\frac{R}{r}\right)^2 \cos\theta . \qquad \text{(S.28)}$$

The external fluid must satisfy the conditions at infinity. In this case, it is necessary to use the solution of the internal problem given by Eqs. (S.23)-(S.25) for $a$=0. Indeed, this will allow us to satisfy conditions (S.1)-(S.2). Substituting Eqs. (S.23)-(S.25) for $a$=0 into Eqs. (S.1)-(S.2) gives

$$2c = V_0 . \qquad \text{(S.29)}$$

Substituting Eq. (S.29) in Eqs. (S.23)-(S.25), one can obtain

$$V_r = \left\{V_0 - b\frac{R}{r} - 2d\left(\frac{R}{r}\right)^3\right\}\cos\theta , \qquad \text{(S.30)}$$

$$V_\theta = \left\{\frac{bR}{2r} - d\left(\frac{R}{r}\right)^3 - V_0\right\}\sin\theta , \qquad \text{(S.31)}$$

$$p = -\frac{\eta}{R} b\left(\frac{R}{r}\right)^2 \cos\theta . \qquad \text{(S.32)}$$

Substituting Eq. (S.30) in Eq. (S.3), we have

$$V_r(R,\theta) = (V_0 - b - 2d)\cos\theta = 0 . \qquad \text{(S.33)}$$

Hence

$$b = V_0 - 2d . \qquad \text{(S.34)}$$

Substituting Eq. (S.34) into Eqs. (S.30)-(S.32), we obtain

$$V_r = \left\{V_0\left(1 - \frac{R}{r}\right) + 2d\left[\frac{R}{r} - \left(\frac{R}{r}\right)^3\right]\right\}\cos\theta , \qquad \text{(S.35)}$$

$$V_\theta = \left\{V_0\left(\frac{R}{2r} - 1\right) - d\left[\frac{R}{r} + \left(\frac{R}{r}\right)^3\right]\right\}\sin\theta , \qquad \text{(S.36)}$$

$$p = -\frac{\eta}{R}(V_0 - 2d)\left(\frac{R}{r}\right)^2 \cos\theta . \qquad \text{(S.37)}$$

Obviously, the internal fluid must be described only by the solution of the internal problem

$$V_r' = 2\left\{\frac{a'}{10}\left(\frac{r}{R}\right)^2 + c'\right\}\cos\theta , \qquad \text{(S.38)}$$

$$V_\theta' = -\left\{a'\frac{4}{10}\left(\frac{r}{R}\right)^2 + 2c'\right\}\sin\theta . \qquad \text{(S.39)}$$

$$p' = p_0' + \frac{2a'\eta' r}{R^2}\cos\theta. \quad (S.40)$$

Taking into account the condition (S.8), for the radial component of the velocity of the internal fluid (S.38) we have

$$V_r'(R,\theta) = 2\left\{\frac{a'}{10} + c'\right\}\cos\theta = 0. \quad (S.41)$$

Hence

$$a' = -10c'. \quad (S.42)$$

Then, the system of equations for the internal fluid (S.38)-(S.40) can be rewritten as

$$V_r' = 2c'\left\{1 - \left(\frac{r}{R}\right)^2\right\}\cos\theta, \quad (S.43)$$

$$V_\theta' = 2c'\left\{2\left(\frac{r}{R}\right)^2 - 1\right\}\sin\theta, \quad (S.44)$$

$$p' = -\frac{20c'\eta' r}{R^2}\cos\theta. \quad (S.45)$$

Conditions given by Eqs. (S.3) and (S.13) imposed in Eq. (S.35) and (S.34) on the surface of the drop give

$$\left.\frac{\partial V_r}{\partial\theta}\right|_{r=R} = \left.\frac{\partial V_r'}{\partial\theta}\right|_{r=R} = 0. \quad (S.46)$$

Therefore, the condition of stitching stresses $\sigma_{r\theta}$ giving by Eq. (S.5) is simplified

$$\eta\left(\left.\frac{\partial V_\theta}{\partial r}\right|_{r=R} - \frac{V_\theta(R)}{R}\right) = \eta'\left(\left.\frac{\partial V_\theta'}{\partial r}\right|_{r=R} - \frac{V_\theta'(R)}{R}\right). \quad (S.47)$$

Next, we have

$$V_\theta(R) = -\left\{\frac{1}{2}V_0 + 2d\right\}\sin\theta, \quad (S.48)$$

$$\frac{\partial V_\theta}{\partial r} = \left\{-V_0\left(\frac{R}{2r^2}\right) + d\left[\frac{R}{r^2} + \frac{3}{R}\left(\frac{R}{r}\right)^4\right]\right\}\sin\theta, \quad (S.49)$$

$$\left.\frac{\partial V_\theta}{\partial r}\right|_{r=R} = \frac{8d - V_0}{2R}\sin\theta, \quad (S.50)$$

$$\left.\frac{\partial V_\theta}{\partial r}\right|_{r=R} - \frac{V_\theta(R)}{R} = \frac{8d - V_0}{2R}\sin\theta + \frac{1}{R}\left\{\frac{1}{2}V_0 + 2d\right\}\sin\theta = \frac{6d}{R}\sin\theta. \quad (S.51)$$

Similarly,

$$V_\theta'(R) = 2c'\sin\theta, \quad (S.52)$$

$$\frac{\partial V_\theta'}{\partial r} = 2c'\frac{1}{R}\left\{4\frac{r}{R}\right\}\sin\theta, \quad (S.53)$$

$$\left.\frac{\partial V_\theta'}{\partial r}\right|_{r=R} = 8c'\frac{1}{R}\sin\theta, \quad (S.54)$$

$$\left.\frac{\partial V_\theta'}{\partial r}\right|_{r=R} - \frac{V_\theta'(R)}{R} = \frac{6c'}{R}\sin\theta, \quad \text{(S.55)}$$

Substituting Eq. (S.51) and Eq. (S.55) into Eq. (S.47), we obtain

$$c' = \frac{\eta}{\eta'}d. \quad \text{(S.56)}$$

Then, the Eqs. (S.43)-(S.45) can be rewritten as

$$V_r' = 2\frac{\eta}{\eta'}d\left\{1-\left(\frac{r}{R}\right)^2\right\}\cos\theta, \quad \text{(S.57)}$$

$$V_\theta' = 2\frac{\eta}{\eta'}d\left\{2\left(\frac{r}{R}\right)^2-1\right\}\sin\theta. \quad \text{(S.58)}$$

$$p' = -\frac{\eta}{\eta'}d\frac{20\eta' r}{R^2}\cos\theta = -d\frac{20\eta r}{R^2}\cos\theta. \quad \text{(S.59)}$$

Let us calculate the expression, taking into account Eqs. (S.35) and (S.37)

$$-p(R) + 2\eta\left.\frac{\partial V_r}{\partial r}\right|_R = \frac{3\eta}{R}(V_0 + 2d)\cos\theta. \quad \text{(S.60)}$$

Similarly, we find

$$-p'(R) + 2\eta'\left.\frac{\partial V_r'}{\partial r}\right|_{r=R} = d\frac{20\eta}{R}\cos\theta - \frac{8\eta'\eta}{\eta' R}d\cos\theta = \frac{12d\eta}{R}\cos\theta. \quad \text{(S.61)}$$

The constancy of the steady-state velocity of the drop is ensured by the resistance force, which depends on the velocity of the drop, is equal in magnitude to the force (S.11) and is directed in the opposite direction. The resistance force is associated with the steady-state flow of the external liquid around the drop and the flow of another liquid inside the drop.
Substituting Eq. (S.60) and Eq. (S.61) into the boundary condition Eq. (S.13), we obtain

$$V_0 = 2d + \frac{(\rho' - \rho)gR^2}{3\eta} \quad \text{(S.62)}$$

The last boundary condition to be taken into account is the partial slip condition for the liquid-liquid interface (S.14). Substituting Eq. (S.46) in Eq. (S.14), we have:

$$\lambda\left(\left.\frac{\partial V_\theta}{\partial r}\right|_{r=R} - \frac{V_\theta(R,\theta)}{R}\right) = V_\theta(R,\theta) - V_\theta'(R,\theta). \quad \text{(S.63)}$$

Substituting into Eq. (S.63) the projections of the velocities $V_\theta$ and $V_\theta'$, determined by Eqs. (S.36) and (S.58), respectively, as well as Eq. (S.51), we find an equation that allows us to calculate the coefficient $d$:

$$\lambda\frac{6d}{R}\sin\theta = \left(-\left\{\frac{1}{2}V_0 + 2d\right\} - 2\frac{\eta}{\eta'}d\right)\sin\theta, \quad \text{(S.64)}$$

or

$$d = -\frac{(\rho' - \rho)gR^2}{6\eta}\left(\frac{6\lambda}{R} + 3 + \frac{2\eta}{\eta'}\right)^{-1}. \quad \text{(S.65)}$$

Substituting Eq. (S.65) into Eq. (S.62), one can obtain

$$V_0 = \frac{(\rho' - \rho) g R^2}{3\eta} \left\{ 1 - \left( \frac{6\lambda}{R} + 3 + \frac{2\eta}{\eta'} \right)^{-1} \right\} \tag{S.66}$$

The remaining undefined coefficients (S.34), (S.56) and (S.42) take the form

$$b = \frac{(\rho' - \rho) g R^2}{3\eta}, \tag{S.67}$$

$$c' = -\frac{(\rho' - \rho) g R^2}{6\eta'} \left( \frac{6\lambda}{R} + 3 + \frac{2\eta}{\eta'} \right)^{-1}, \tag{S.68}$$

$$a' = \frac{5(\rho' - \rho) g R^2}{3\eta'} \left( \frac{6\lambda}{R} + 3 + \frac{2\eta}{\eta'} \right)^{-1}. \tag{S.69}$$

Let us write out the solutions for the external and internal liquids by substituting Eq. (S.65) in Eqs. (S.35)-(S.37) and Eqs. (S.57)-(S.59), respectively.

1) The external liquid

$$V_r = \frac{(\rho' - \rho) g R^2}{3\eta} \left\{ 1 - \frac{R}{r} - \left( \frac{6\lambda}{R} + 3 + \frac{2\eta}{\eta'} \right)^{-1} \left[ 1 - \left( \frac{R}{r} \right)^3 \right] \right\} \cos\theta, \tag{S.70}$$

$$V_\theta = \frac{(\rho' - \rho) g R^2}{6\eta} \left\{ \frac{R}{r} - 2 + \left( \frac{6\lambda}{R} + 3 + \frac{2\eta}{\eta'} \right)^{-1} \left[ 2 + \left( \frac{R}{r} \right)^3 \right] \right\} \sin\theta, \tag{S.71}$$

$$p = -\frac{(\rho' - \rho) g R^3}{3r^2} \cos\theta. \tag{S.72}$$

Stream function is (Table S1):

$$\Psi(r, \theta) = \frac{(\rho' - \rho) g R^4}{6\eta} \sin^2\theta \left\{ \left( \frac{r}{R} \right)^2 - \frac{r}{R} + \left( \frac{6\lambda}{R} + 3 + \frac{2\eta}{\eta'} \right)^{-1} \left( \frac{R}{r} - \left( \frac{r}{R} \right)^2 \right) \right\}. \tag{S.73}$$

2) The internal liquid

$$V_r' = -\frac{(\rho' - \rho) g R^2}{3\eta'} \left( \frac{6\lambda}{R} + 3 + \frac{2\eta}{\eta'} \right)^{-1} \left\{ 1 - \left( \frac{r}{R} \right)^2 \right\} \cos\theta, \tag{S.74}$$

$$V_\theta' = -\frac{(\rho' - \rho) g R^2}{3\eta'} \left( \frac{6\lambda}{R} + 3 + \frac{2\eta}{\eta'} \right)^{-1} \left\{ 2\left( \frac{r}{R} \right)^2 - 1 \right\} \sin\theta, \tag{S.75}$$

$$p' = \frac{10(\rho' - \rho) g r}{3} \left( \frac{6\lambda}{R} + 3 + \frac{2\eta}{\eta'} \right)^{-1} \cos\theta. \tag{S.76}$$

Stream function is (Table S1):

$$\Psi' = \frac{(\rho' - \rho) g R^4}{6\eta'} \sin^2\theta \left( \frac{6\lambda}{R} + 3 + \frac{2\eta}{\eta'} \right)^{-1} \left\{ \left( \frac{r}{R} \right)^4 - \left( \frac{r}{R} \right)^2 \right\}. \tag{S.77}$$

Eq. (S.66) describes the steady-state velocity of a spherical liquid drop in an external fluid. It can also be represented in the equivalent form

$$V_0 = \frac{2(\rho' - \rho) g R^2}{3\eta} \frac{1 + \eta\eta'^{-1} + 3\lambda R^{-1}}{3 + 2\eta\eta'^{-1} + 6\lambda R^{-1}} \tag{S.78}$$

or

$$V_0 = \frac{2(\rho'-\rho)gR^2}{3\eta}\frac{\eta+\eta'(1+3\lambda R^{-1})}{2\eta+3\eta'(1+2\lambda R^{-1})}, \tag{S.79}$$

or

$$V_0 = \frac{2(\rho'-\rho)gR^2}{3\eta}\frac{\alpha+1+3\lambda R^{-1}}{2\alpha+3+6\lambda R^{-1}}, \tag{S.80}$$

where $\alpha = \frac{\eta}{\eta'}$ is the viscosity ratio. Note that Eq. (S.80) can also be represented as

$$V_0 = \frac{(\rho'-\rho)gR^2}{3\eta}\left\{1-\left(\frac{6\lambda}{R}+3+2\alpha\right)^{-1}\right\} \tag{S.81}$$

At $\lambda=0$, we have the no-slip condition, under which Eq. (S.79) becomes the Hadamard–Rybczynski equation (HRE):

$$V_0 = \frac{2(\rho'-\rho)gR^2}{3\eta}\frac{\eta+\eta'}{2\eta+3\eta'}. \tag{S.82}$$

At $\lambda=0$, condition given by Eq. (S.57) takes the form

$$V_\theta(R,\theta) - V_\theta'(R,\theta) = 0. \tag{S.83}$$

It is no-slip condition that corresponds to the Hadamard–Rybczynski model. Thus, the resulting Eq. (S.66) generalizes the Hadamard–Rybczynski equation and its limit is the Stokes formula by introducing partial slip.

If $\alpha \to 0$, i.e. the viscosity of the drop is much greater than the viscosity of the surrounding liquid, and $\lambda$ has a value limited in absolute value from above, Eq. (S.80) takes the form

$$V_0 = \frac{2(\rho'-\rho)gR^2}{9\eta}\frac{1+3\lambda R^{-1}}{1+2\lambda R^{-1}}, \tag{S.84}$$

Eq. (S.84) describes the sliding of a solid sphere in an external fluid with viscosity $\eta$ and slip length $\lambda$. It coincides with the previously published result, Ref. [S4,S5]. At $\lambda=0$, Eq. (S.84) becomes the Stokes equation that describes the motion of a solid sphere under the no-slip boundary condition:

$$V_0 = \frac{2(\rho'-\rho)gR^2}{9\eta}. \tag{S.85}$$

Note that, taking into account Eq. (S.16), we have

$$\alpha = -\frac{\lambda}{\lambda'}. \tag{S.86}$$

Then, the original formula (S.80) transforms to the Stokes equation in the limit $\lambda \to 0$, when simultaneously $\lambda R^{-1} \to 0$ and $\lambda(\lambda')^{-1} \to 0$, i.e., when the external liquid satisfies to a condition close to no-slip, but at the same time, the partial slip regime must take place in the internal liquid. This means that the Stokes limit (S.85) can be fulfilled not only for a solid sphere moving in the no-slip mode, as was previously assumed, but also under the condition of partial slip of the internal liquid, which is assumed to be much more viscous than the external one.

Another extreme case is when the viscosity of the external liquid is much higher than the viscosity of the drop. If $\lambda$ is a limited quantity, the substitution $\alpha \to \infty$ transforms Eq. (S.80) into the following expression

$$V_0 = \frac{(\rho' - \rho) g R^2}{3\eta}. \quad \text{(S.87)}$$

The HRE has the same limit.

Note that Eq. (S.80) gives another way to obtain Eq. (S.87), which is absent in the HRE: i.e., let $\lambda R^{-1} \to \infty$ and, at the same time, α is a limited value. This means that the slip of the external liquid is almost complete, and that of the internal liquid is the same, taking into account Eq. (S.86).

Considering that the force acting on a liquid drop in an external liquid is determined by the formula

$$F_0 = \frac{4\pi}{3}(\rho' - \rho) g R^3. \quad \text{(S.88)}$$

Then the friction coefficient is determined by the formula

$$\beta = \frac{F_0}{V_0} = 2\pi R\eta \frac{2\alpha + 3 + 6\lambda R^{-1}}{\alpha + 1 + 3\lambda R^{-1}}. \quad \text{(S.89)}$$

In the limit $\lambda R^{-1} \to 0$ and $\alpha \to 0$ that corresponds to a solid body with the boundary condition no-slip, Eq. (S.89) takes the form $\beta = 6\pi R\eta$, which coincides with the Stokes friction coefficient, as it should be. In the case of complete slip, i.e., $\lambda R^{-1} \to \infty$, on can obtain $\beta = 4\pi R\eta$.

We now obtain an expression for the frequently used drag coefficient $C_D$, Ref. [S.6], that is determined according to the relation

$$C_D = \frac{2F_0}{\pi R^2 \rho V_0^2} = \frac{2\beta}{\pi R^2 \rho V_0}. \quad \text{(S.90)}$$

Using Eq. (S.89) and the Reynolds number

$$\mathrm{Re} = \frac{2\rho R V_0}{\eta}, \quad \text{(S.91)}$$

rewrite Eq. (S.90) as

$$C_D = \frac{4\beta}{\pi R\eta \,\mathrm{Re}} = \frac{8}{\mathrm{Re}} \frac{2\alpha + 3 + 6\lambda R^{-1}}{\alpha + 1 + 3\lambda R^{-1}}. \quad \text{(S.92)}$$

In the Hadamard–Rybczynski limit, i.e., if $\lambda R^{-1} \to 0$, we obtain the well-known expression, Ref. [S7]

$$C_{D,AR} = \frac{8}{\mathrm{Re}} \frac{2\alpha + 3}{\alpha + 1}. \quad \text{(S.93)}$$

In the case of complete slip, i.e., if $\lambda R^{-1} \to \infty$, one can obtain

$$C_D = \frac{16}{\mathrm{Re}}. \quad \text{(S.94)}$$

### 3. Gas-liquid interface and small bubble rise

Let us consider the motion of a gas bubble. In this case, the normal slip condition (S.22) should be used instead of condition (S.13). Taking into account Eq. (S.22), for the radial component of the velocity of the internal fluid (S.38) we have

$$V_r'(R,\theta) = 2\left\{\frac{a'}{10} + c'\right\}\cos\theta = \frac{Df_1}{n_0 R}\cos\theta. \quad \text{(S.95)}$$

Hence

$$a' = \frac{5Df_1}{n_0 R} - 10c'. \tag{S.96}$$

Then, the system of equations for the internal fluid (S.38)-(S.40) can be rewritten as

$$V_r' = \frac{Df_1}{n_0 R}\left(\frac{r}{R}\right)^2 \cos\theta + 2c'\left\{1 - \left(\frac{r}{R}\right)^2\right\}\cos\theta, \tag{S.97}$$

$$V_\theta' = -\frac{2Df_1}{n_0 R}\left(\frac{r}{R}\right)^2 \sin\theta + 2c'\left\{2\left(\frac{r}{R}\right)^2 - 1\right\}\sin\theta. \tag{S.98}$$

$$p' = p_0' + 10\left(\frac{Df_1}{n_0 R} - 2c'\right)\frac{\eta' r}{R^2}\cos\theta. \tag{S.99}$$

Eqs. (S.97)-(S.99) on the surface of the drop gives

$$\left.\frac{\partial V_r'}{\partial\theta}\right|_{r=R} = -\frac{Df_1}{n_0 R}\sin\theta, \tag{S.100}$$

$$\left.\frac{\partial V_\theta'}{\partial r}\right|_{r=R} = -\frac{4Df_1}{n_0 R^2}\sin\theta + \frac{8c'}{R}\sin\theta, \tag{S.101}$$

$$\frac{V_\theta'(R,\theta)}{R} = -\frac{2Df_1}{n_0 R^2}\sin\theta + \frac{2c'}{R}\sin\theta. \tag{S.102}$$

Then $\sigma_{r\theta}'$ takes the form

$$\sigma_{r\theta}' = \eta'\left(\frac{1}{R}\frac{\partial V_r'}{\partial\theta} + \frac{\partial V_\theta'}{\partial r} - \frac{V_\theta'}{R}\right)_{r=R} = \eta'\left(\frac{6c'}{R} - \frac{3Df_1}{n_0 R^2}\right)\sin\theta. \tag{S.103}$$

Substituting (S.103) and (S.51) into (S.4), we obtain

$$c' = \frac{\eta}{\eta'}d + \frac{Df_1}{2n_0 R}. \tag{S.104}$$

Then, the Eqs. (S.97)-(S.99) can be rewritten as

$$V_r' = \frac{Df_1}{n_0 R}\left(\frac{r}{R}\right)^2 \cos\theta + 2\left(\frac{\eta}{\eta'}d + \frac{Df_1}{2n_0 R}\right)\left\{1 - \left(\frac{r}{R}\right)^2\right\}\cos\theta, \tag{S.105}$$

$$V_\theta' = -\frac{2Df_1}{n_0 R}\left(\frac{r}{R}\right)^2 \sin\theta + 2\left(\frac{\eta}{\eta'}d + \frac{Df_1}{2n_0 R}\right)\left\{2\left(\frac{r}{R}\right)^2 - 1\right\}\sin\theta. \tag{S.106}$$

$$p' = p_0' - 20\frac{\eta d r}{R^2}\cos\theta. \tag{S.107}$$

Calculating the derivative on the interface

$$\left.\frac{\partial V_r'}{\partial r}\right|_{r=R} = -4\frac{\eta}{\eta' R}d\cos\theta, \tag{S.108}$$

we find

$$-p'(R) + 2\eta'\left.\frac{\partial V_r'}{\partial r}\right|_{r=R} = d\frac{20\eta}{R}\cos\theta - \frac{8\eta'\eta}{\eta' R}d\cos\theta = \frac{12d\eta}{R}\cos\theta. \tag{S.109}$$

Substituting Eq. (S.60) and Eq. (S.109) into the boundary condition Eq. (S.12), we obtain

$$V_0 = 2d + \frac{(\rho' - \rho)gR^2}{3\eta} \tag{S.110}$$

Let us use the longitudinal slip condition (S.17)

$$\left[-p' + 2\eta' \frac{\partial V_z'}{\partial r}\right]_{r=R} = \left[-k_B T \Delta n + \frac{k_B T \lambda'}{\pi} \frac{\partial n}{\partial r}\right]_{r=R}, \tag{S.111}$$

which, when substituting (S.20), takes the form

$$\left[-p' + 2\eta' \frac{\partial V_z'}{\partial r}\right]_{r=R} = -k_B T f_1 \left[1 - \frac{\lambda'}{\pi R}\right], \tag{S.112}$$

or, taking into account (S.109),

$$\frac{12 d\eta}{R} = -k_B T f_1 \left[1 - \frac{\lambda'}{\pi R}\right], \tag{S.112}$$

we find

$$f_1 = -\frac{12 d\eta}{R k_B T}\left[1 - \frac{\lambda'}{\pi R}\right]^{-1}. \tag{S.113}$$

Finally, we use the transverse slip condition (S.63), taking into account the following formulae

$$V_\theta'(R) = -\frac{D f_1}{n_0 R}\sin\theta + \frac{2\eta}{\eta'} d \sin\theta, \tag{S.114}$$

$$V_\theta(R,\theta) = \left\{-\frac{1}{2}V_0 - 2d\right\}\sin\theta. \tag{S.115}$$

Taking into account (S.51), one can obtain

$$\lambda \frac{6d}{R}\sin\theta = \left(-\left\{\frac{1}{2}V_0 + 2d\right\} - 2\frac{\eta}{\eta'} d + \frac{D f_1}{n_0 R}\right)\sin\theta. \tag{S.116}$$

Substituting (S.113) into (S.116), we find

$$d = -\frac{1}{4}V_0 \left\{\frac{3\lambda}{R} + 1 + \frac{\eta}{\eta'} + \frac{3\lambda'^2}{\pi R^2}\frac{\eta}{\eta'}\left[1 - \frac{\lambda'}{\pi R}\right]^{-1}\right\}^{-1}. \tag{S.117}$$

Substituting (S.117) into (S.110), we arrive at the expression for the bubble's rise velocity

$$V_0 = \frac{2(\rho' - \rho)gR^2}{3\eta}\left\{\frac{3\lambda}{R} + 1 + \frac{\eta}{\eta'} + \frac{3\lambda'^2}{\pi R^2}\frac{\eta}{\eta'}\left[1 - \frac{\lambda'}{\pi R}\right]^{-1}\right\}\left\{\frac{6\lambda}{R} + 3 + \frac{2\eta}{\eta'} + \frac{6\lambda'^2}{\pi R^2}\frac{\eta}{\eta'}\left[1 - \frac{\lambda'}{\pi R}\right]^{-1}\right\}^{-1} \tag{S.118}$$

In formula (S.118), the terms containing the factor $\lambda'^2$ are responsible for the longitudinal slip. Eliminating these terms, we again arrive at formula (S.66) or, equivalently, (S.78).

Up to the second power, in the expansion of the numerator and denominator of (S.118) in $\lambda'$, we have

$$V_0 = \frac{2(\rho' - \rho)gR^2}{3\eta}\left\{\frac{3\lambda}{R} + 1 + \frac{\eta}{\eta'}\left(1 + \frac{3\lambda'^2}{\pi R^2}\right)\right\}\left\{\frac{6\lambda}{R} + 3 + \frac{2\eta}{\eta'}\left(1 + \frac{3\lambda'^2}{\pi R^2}\right)\right\}^{-1} \tag{S.119}$$

or

$$V_0 = \frac{(\rho' - \rho)gR^2}{3\eta}\left\{\frac{3|\lambda'|}{R} + \frac{\eta'}{\eta} + 1 + \frac{3\lambda'^2}{\pi R^2}\right\}\left\{\frac{3|\lambda'|}{R} + \frac{3\eta'}{2\eta} + 1 + \frac{3\lambda'^2}{\pi R^2}\right\}^{-1}, \tag{S.120}$$

where the slip length in the bubble $\lambda'$ is determined by formula (S.116).
Substituting (S.119) into (S.113), in the zeroth order of smallness in the expansion in powers of $\lambda' R^{-1}$, we find

$$\frac{f_1}{n_0} \approx -\frac{\rho g R}{p_a}, \tag{S.121}$$

where $p_a = n_0 k_B T$ is an atmospheric pressure.
Taking into account (S.20), for the gas density inside the bubble we have

$$n = n_0 \left(1 - \frac{\rho g r}{p_a} \cos\theta\right), \tag{S.122}$$

Thus, the bubble density is slightly lower at the top and slightly higher at the bottom than at the center (Fig.S1). However, at the Reynolds numbers considered, this effect is very small.

The characteristic relative increment in air density $\frac{\rho g R}{p_a}$ for a bubble of radius $R = 10\,\mu\text{m}$ is $10^{-6}$. This value increases with increasing bubble radius, and as the radius approaches 1 mm, it can reach a value that leads to noticeable deformation of the bubble. However, the rise regime of such large bubble meets the criteria Re>1 and Pe>1, which exceeds the limitations adopted here and requires taking into account higher-rank spherical modes and the corresponding shape deformations.

**References to the Supplementary**